\documentclass[letterpaper,usenatbib]{mn2e}
\usepackage{graphicx}
\usepackage{color}
\usepackage{lscape}
\usepackage{rotating}
\usepackage{longtable}
\usepackage{epstopdf}
\newsavebox{\tablebox}
\usepackage{hyperref}
\usepackage{times}
\usepackage{longtable,lscape}

\newcommand{\ergs}{\ifmmode {\rm erg\ s}^{-1} \else erg s$^{-1}$\ \fi}

\newcommand{\feii}{Fe {\sc ii}\ }
\newcommand{\mgii}{Mg {\sc ii}\ }
\newcommand{\civ}{C {\sc iv}\ }
\newcommand{\ciii}{C {\sc iii]}\ }

\newcommand{\siiv}{Si {\sc iv}\ }

\newcommand{\lb}{\ifmmode L_{\rm bol} \else $L_{\rm bol}$\ \fi}
\newcommand{\ledd}{\ifmmode L_{\rm Edd} \else $L_{\rm Edd}$\ \fi}
\newcommand{\leddR}{\ifmmode L_{\rm bol}/L_{\rm Edd} \else $L_{\rm bol}/L_{\rm Edd}$\ \fi}
\newcommand{\lx}{\ifmmode L_{\rm 2-10keV} \else  $L_{\rm 2-10keV}$\ \fi}
\newcommand{\hb}{\ifmmode H\beta \else H$\beta$\ \fi}
\newcommand{\ha}{\ifmmode H\alpha \else H$\alpha$\ \fi}
\newcommand{\hg}{\ifmmode H\alpha \else H$\gamma$\ \fi}

\newcommand{\mbh}{\ifmmode M_{\rm BH}  \else $M_{\rm BH}$\ \fi}
\newcommand{\lv}{\ifmmode \lambda L_{\lambda}(1350\AA) \else $\lambda L_{\lambda}(1350\AA)$\ \fi}
\newcommand{\lcon}{\ifmmode L_{1350} \else $L_{1350}$\ \fi}

\newcommand{\mdot}{\ifmmode \dot{m} \else \dot{m} \fi }
\newcommand{\llog}{\ifmmode {\rm log} \else {\rm log} \fi }
\newcommand{\kms}{\ifmmode {\rm km\ s}^{-1} \else km s$^{-1}$\ \fi}

\begin{document}
\title[BAL variable regions]{Variability of QSOs with variable regions in broad absorption troughs from the Sloan Digital Sky Survey}
\author[Z.-C. He et al.]{Zhi-Cheng He,
Wei-Hao Bian\thanks{E-mail: whbian@njnu.edu.cn}, Xiao-Lei Jiang and Xue Ge \\
Department of Physics and Institute of Theoretical Physics, Nanjing
Normal University, Nanjing 210023, China\\} \maketitle

\begin{abstract}
The variability of broad absorption lines is investigated for a sample of 188 broad-absorption-line (BAL) quasars (QSOs) ($z > 1.7$) with at least two-epoch observations from the Sloan Digital Sky Survey Data Release 7 (SDSS DR7), covering a time-scale of about 0.001 -- 3 years in the rest frame. Considering only the longest time-scale between epochs for each QSO, 73 variable regions in the \civ BAL troughs are detected for 43 BAL QSOs. The proportion of BAL QSOs showing variable regions increases with longer time-interval than about 1 year in the rest frame. The velocity width of variable regions is narrow compared to the BAL-trough outflow velocity. For 43 BAL QSOs with variable regions, it is found that there is a medium strong correlation between the variation of the continuum luminosity at 1500 \AA\ and the variation of the spectral index. With respect to the total 188 QSOs, larger proportion of BAL QSOs with variable regions appears bluer during their brighter phases, which implies that the origin of BAL variable regions is related to the central accretion process. For 43 BAL QSOs with variable regions, it is possible that there is a negative medium strong correlation between the absolute variation of the equivalent width and the \mgii-based black hole mass, and a medium strong correlation between the maximum outflow velocity of variable regions and the Eddington ratio. These results imply the connection between the BAL-trough variation and the central accretion process.


\end{abstract}


\section{INTRODUCTION}
Broad absorption line quasars (BAL QSOs) exhibit broad absorption troughs for high-ionization ultraviolet (UV) lines such as \siiv 1399, \civ 1549, \ciii 1909 (known as HiBAL), or/and low-ionization UV lines such as \mgii 2799 (known as LoBAL). BAL QSOs were usually identified from their UV spectra by the balnicity index (BI), considering different BAL outflow velocity and/or BAL velocity width \citep[e.g.][]{Weymann91, Trump06, Gibson09, He14}. BAL troughs are present in about 10-40\% of QSOs \citep[e.g.][] {Gibson09, Allen11}.

BAL troughs in QSOs are thought to be the strongest observed signature of QSO winds \citep{Fabian12}. One explanation for BALs is an orientation-dependent effect, where QSOs appear as BAL QSOs when the disk wind is on the line of sight \citep[e.g.][]{Murray95, Elvis00}. The requirement that detection of a wind should be orientation-dependent is very similar to the case of other structures in QSOs, such as the broad-line region (BLR) or jet (e.g. Urry \& Padovani 1995). Another explanation for BALs is an orientation-independent evolution effect \citep[e.g.][]{Gibson08, Zubovas13}. As an evolutionary stage of active galactic nucleus (AGN), the expulsion of gas and dust by galaxy collision possibly causes the BAL outflows. It was supported by no correlations exist between outflow properties and orientation, such as the similar range of viewing angles for radio-loud BALs and radio-loud non-BALs \citep[e.g.][]{Fine11, Bruni12}.

The disk wind in BAL QSOs is believed to come from the central supermassive black hole (SMBH) accretion disc,  and BAL region often lies outside the BLR region \citep[e.g.][]{Proga00, Murray95,Filiz13}. There exists an empirical relation between the BLR size and the continuum luminosity in AGN/QSOs \citep[e.g.][]{Kaspi05,Bentz09}. For the BAL region outside of BLR region, larger luminosity would lead to larger size, smaller orbital velocity of the BAL regions. It would lead the properties/variability of BAL outflows as a function of luminosity. The wind dependence on the properties of QSOs has been discussed by many authors, suggesting the connection between the outflow and the accretion process \citep[e.g.][]{Laor02, Ganguly07, Baskin13, He14}. It was found that the maximum outflow velocity increases with both the bolometric luminosity and the blueness of the spectral slope, suggesting the idea of radiation-pressure-driven outflows \citep{Laor02, Ganguly07}. At the same time, BAL-troughs often vary over rest-frame time-scales of days to years, which can provide clues to the origin of BAL QSOs \citep[e.g.][]{Gibson08, Capellupo11, Capellupo12, Capellupo13, Filiz12, Filiz13, Grier15}. \cite{Capellupo11, Capellupo12, Capellupo13} investigated BAL variability for 24 QSOs with long time-scales and multiple epochs. They found that variability typically occurs only in portions of the BAL troughs; the components at higher outflow velocities are more likely to vary than those at lower velocities and weaker BALs are more likely to vary than stronger BALs; both the incidence and the amplitude of variability are greater across longer time-scales. With a sample of 291 BAL QSOs from SDSS-I/II/III, \cite{Filiz13} investigated BAL variations as a function of QSOs properties and did not find significant evidence for correlations between BAL variability and luminosity/Eddington ratio/SMBH mass. For \civ 1549 troughs on moderate time-scales (1-2.5 yrs), they suggested possible correlations between BAL variability and luminosity/Eddington ratio. \cite{Filiz13} used \civ emission line to calculate the SMBH mass and the Eddington ratio, which have larger scatters and bias compared to that by other emission lines such as \hb, \mgii.

For some samples of QSOs, it has been shown that spectra of QSOs at low redshifts are bluer during their brighter phases \citep[e.g.][]{Wilhite05, Pu06, Meusinger11, Zuo12, Bian12a}. The trend of bluer spectra during brighter phases is usually explained by the accretion variation. During the brighter phase, the accretion disk becomes hotter and its emission peak would move to shorter wavelengths (big blue bump), which would also lead to larger variance in the blue spectrum. However, other contributions may mingle in this kind of investigation, such as the contribution of UV/optical \feii, Balmer continuum, jet, and the host, as well as the complex of accretion disk model. With two-epoch variation, it was found that the spectra of half of the QSOs appear redder during their brighter phases \citep{Bian12a, Guo14}.

The feature of variable regions in BAL-trough were investigated by some authors \citep[e.g.][]{Gibson08, Filiz12, Filiz13}. Using a sample of 13 BAL QSOs with two-epoch spectra covering 3-6 years in the rest frame, \cite{Gibson08} identified the variable regions in BAL-trough and investigated the variation of \civ BAL-trough between the Large Bright Quasar Survey \citep[LBQS;][]{Hewett95} and the Sloan Digital Sky Survey (SDSS). 
With SDSS/Baryon Oscillation Spectroscopic Survey (BOSS), we have investigated the relation between the wind and QSO properties for a single BAL QSO with 18 epochs observations covering about 3 years in the rest frame \citep{He14}. Here, we present a sample of 188 BAL QSOs with at least two-epoch observation from SDSS Data Release 7 (DR7). This sample is used to investigate the feature of the BAL-trough variable regions, the variability of UV spectral index and the BAL-trough equivalent width (EW), as well as the relation with the central accretion properties. Section 2 presents the sample. Section 3 gives the spectral analysis. Section 4 contains our results and discussion. A summery is given in the last section. Throughout this work, we use a cosmology with $H_0 = 70 \kms \rm Mpc^{-1}$, $\Omega_M = 0.3$, and $\Omega_{\Lambda} = 0.7$.

\section{The sample of \civ BAL QSOs with two-epoch observations from the SDSS DR7}

\begin{figure}
\begin{center}
\includegraphics[height=8cm,angle=-90]{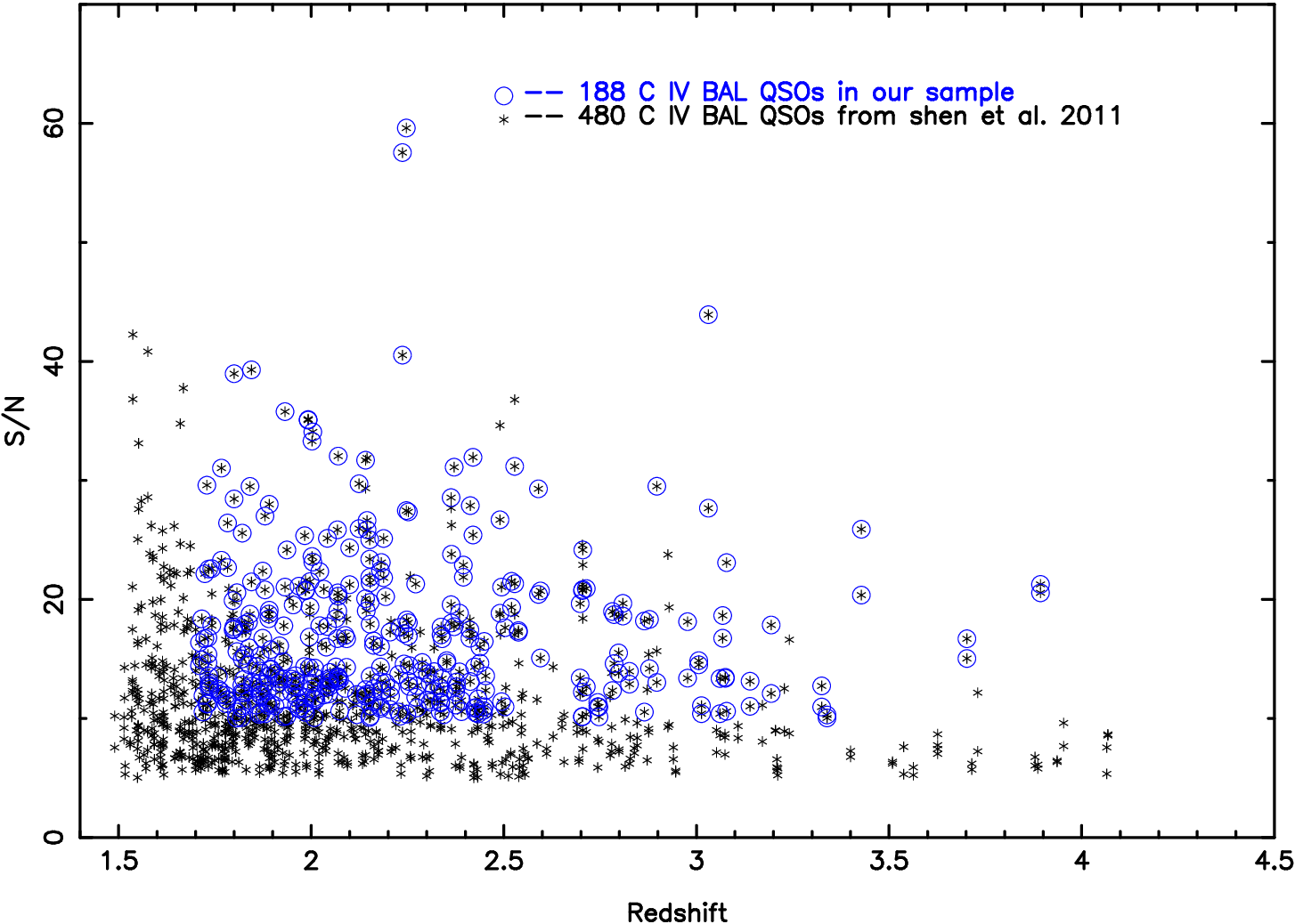}
\includegraphics[height=8cm,angle=-90]{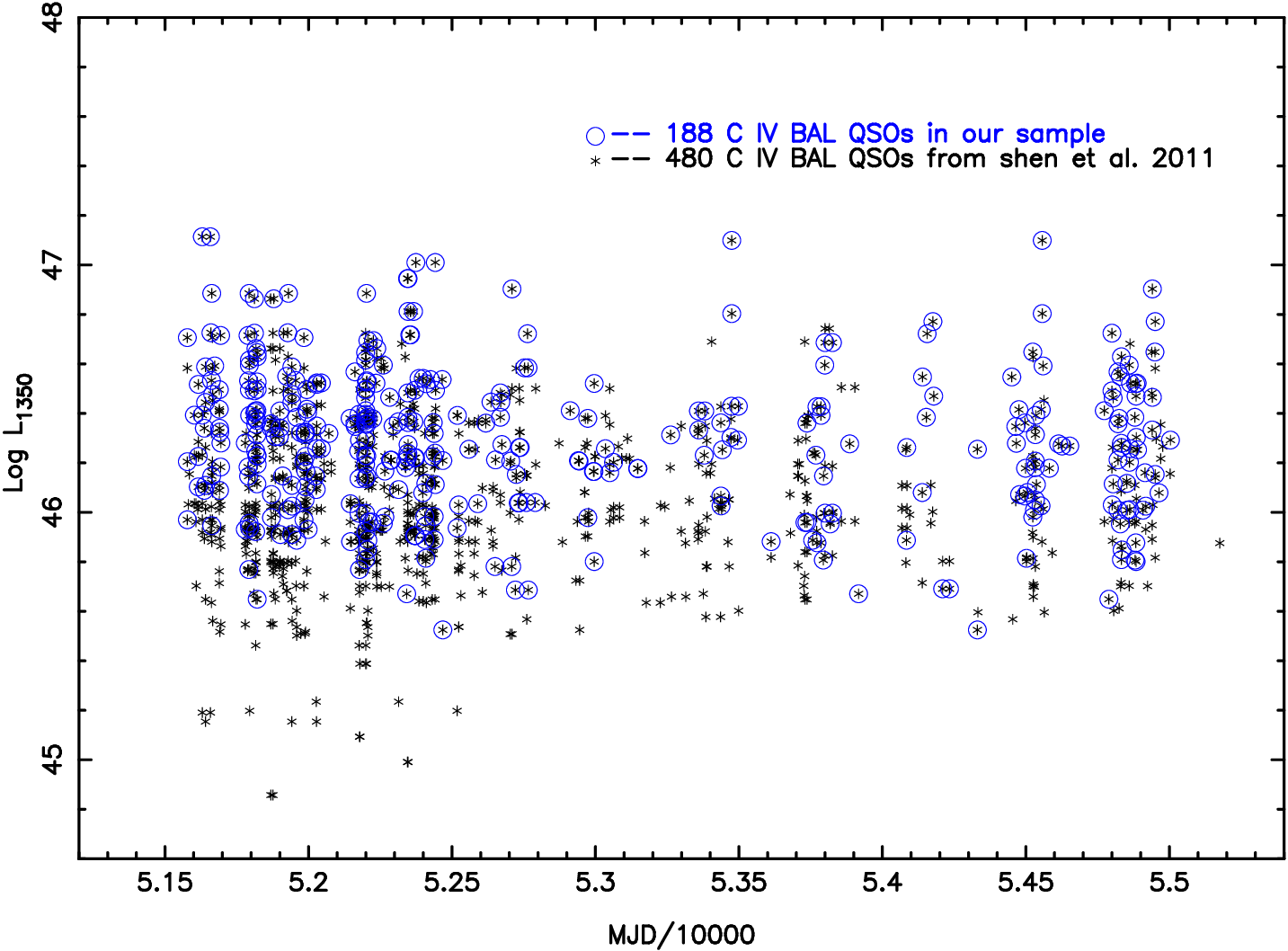}
\caption{Top: the S/N at r-band versus the redshift. Bottom: the continuum luminosity at 1350 \AA\ ($L_{1350}$, in units of \ergs) versus the MJD. The black stars denote 480 BAL QSOs with at least two-epoch spectra from SDSS DR7. The blue circles denote 188 BAL QSOs with longest time-scale two-epoch spectra with S/N(r) $>$ 10 and z$>$1.7. }
\label{fig1}
\end{center}
\end{figure}

The SDSS DR7 \citep{York00} contains imaging of almost 11663 $\rm deg^2$ and spectra for roughly $93\times 10^4$ galaxies and $12\times 10^4$ QSOs, observed by 2.5 m telescope at the Apache Point Observatory in New Mexico. The SDSS spectra were obtained through 3" fibers. For the SDSS DR7 spectra, the observational wavelength coverage is from 3800 \AA\ to 9200 \AA, and the spectral resolution is $1850-2200$. With SDSS DR7, \cite{Shen11} gave a compilation of properties of $105783$ QSOs. With the modified balnicity index $\rm BI_0$ ($0~\kms$ as a minimum detection limit for the outflow velocity range), there are 6214 BAL QSOs identified in their SDSS DR7 sample. It is found that there are 1080 spectra of 480 \civ BAL QSOs with at least two-epoch observations. In order to identify the variable regions in \civ BAL troughs, we select BAL QSOs with SDSS spectral signal-to-noise ratio (S/N) at r band larger than 10, and redshift more than 1.7. It is consist of 188 \civ BAL QSOs with 428 SDSS spectra. For each QSO with more than two epochs, we consider only the longest time-scale between two epochs. In Fig. \ref{fig1}, we give the S/N versus the redshift, and the luminosity at 1350\AA\ ($L_{1350}$) versus the spectral MJD values. The stars denote 480 \civ BAL QSOs with at least two-epoch observations, and the blue circles denote our 188 \civ BAL QSOs with longest two-epoch SDSS spectra with S/N(r) $>$ 10 and z$>$1.7. In Fig. \ref{fig1}, for our sample of 188 BAL QSOs, the redshift distribution is between 1.7 and about 3.9, the S/N distribution is between 10 and about 60. Considering the problem of spectrophotometric flux calibration for BOSS in SDSS III, we just use the spectra from SDSS DR7 \citep[e.g.][]{Paris14, Margala15}. Our sample is different to that by \cite{Filiz13}, who used two-epoch spectra from the SDSS and the BOSS respectively.

\section{Spectral Analysis}

\subsection{Fitting the spectral continuum}

\begin{figure}
\begin{center}
\includegraphics[height=8cm,angle=-90]{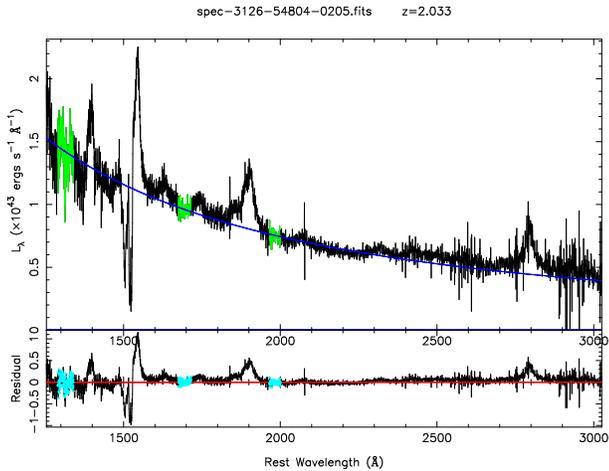}
\caption{An example of the power-law continuum fit. Top: the extinction-corrected rest-frame spectrum is shown as the black line. The green dots are the initial continuum windows. The blue line is the power-law continuum. Bottom: the residual spectrum.}
\label{fig2}
\end{center}
\end{figure}

\begin{figure}
\begin{center}
\includegraphics[height=8cm,angle=-90]{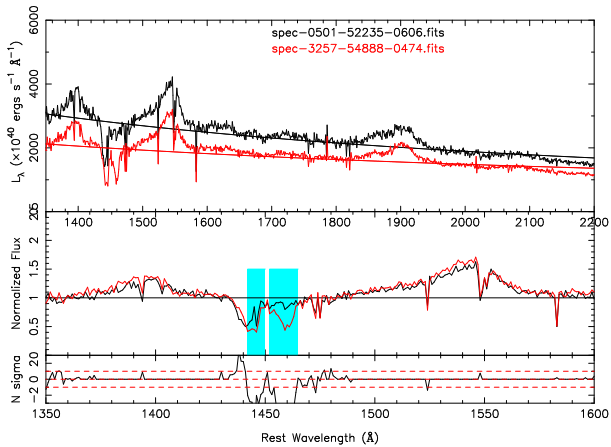}
\caption{An example of the \civ BAL-trough variable regions. Top: two-epoch spectra as well as their continuums. Middle: their normalized spectra and the variable regions in \civ BAL troughs (cyan shaded regions). Bottom: the $N_\sigma$ versus the wavelength $\lambda$. Variable regions of \civ BAL troughs are identified to be where an absorption feature is detected with $|N_{\sigma}| \geq 1$ for at least five consecutive data points (4 \AA\ wide). }
\label{fig3}
\end{center}
\end{figure}

In order to characterize the spectral variation of BAL QSOs and the variable regions in the \civ BAL troughs, we fit the continuum spectrum by a power-law function iteratively \citep{He14}. All the SDSS spectra are corrected for Galactic extinction, assuming the extinction curve of \cite{Cardelli89} (IR band; UV band) and \cite{ODonnell94} (optical band) with $R_V = 3.1$. The $A_V$ values of these SDSS BAL QSOs are derived from the $SpecPhotoAll$ table in SDSS. All spectra are then corrected to the rest frame, and rebinned to a uniform grid with 1\AA\ wide. It is popularly accepted that we can use one power-law function $f_\lambda \propto \lambda ^{\alpha} $ ($f_\nu \propto \nu ^{-(2+\alpha)} $) to fit the QSO continuum. Some authors also used two power-law functions to model the QSO continuum \citep[e.g.][]{Forster01, Shang05}. For BAL QSOs, a polynomial function or a reddened power law was also used to fit the continuum in previous studies \citep{Lundgren07, Gibson08, Gibson09}. Here, we use one power-law function, $f_{\lambda}= f_{2000}(\lambda/2000)^{\alpha}$, to fit the continuum spectra of BAL QSOs \citep{Hu08, He14}. The power-law continuum is fit iteratively in the "continuum windows", which are known to be relatively free from strong emission lines. Our adopted continuum windows are 1290-1330, 1685-1715, 1970-2010 \AA\ in the rest frame \citep[Fig. \ref{fig2}, e.g.][]{Forster01, Vanden01, Gibson08, Bian12a, Baskin13}. The fit is achieved by minimizing $\chi^2$ iteratively. At each iteration, we neglect any spectral bins that deviate by more than $3 \sigma$ from the previous continuum fit. This iterative method would automatically adjust the fitting window, excluding spectral regions that contain broad emission or absorption features, especially for BAL QSOs.

For the power-law continuum, $f_{\lambda}= f_{2000}(\lambda/2000)^{\alpha}$, the error
of the power-law continuum by error propagation is calculated as follows \citep{He14}:
\begin{equation}
 \delta(f_{con})=f_{\lambda} \sqrt{\left(\frac{\delta(f_{2000})}{f_{2000}}\right)^{2}+(\ln\lambda-\ln2000)^{2}\delta\alpha^{2}}.
\label{eq1}
 \end{equation}
where the errors ($\delta(f_{2000})$ and $\delta \alpha$) are given in the power-law fitting. The total \civ BAL-trough EW is calculated as follows:
\begin{equation}
EW =\int_{}^{} [ 1-\frac {f_{obv}(\lambda)}{f_{con}(\lambda)} ] d\lambda.
\label{eq2}
\end{equation}
The integration is integrated from the total \civ BAL-trough for $f_{obv}(\lambda) < f_{con}(\lambda)$.
And the error for the \civ BAL EW is measured as follows:
\begin{equation}
\delta(EW)= \sqrt{\sum_{\lambda}{\left(\frac{f_{obv}}{f_{con}}\right)}^{2}
\left[\left(\frac{\delta(f_{obv})}{f_{obv}}\right)^{2}+\left(\frac{\delta(f_{con})}{f_{con}}\right)^{2}\right]}.
\label{eq3}
\end{equation}
where $\delta(f_{obv})$ at $\lambda $ is the flux error for the SDSS spectrum.

With the continuum fit, we can obtain the spectral index $\alpha$, the continuum luminosity at 1500\AA, and the total \civ BAL EW, as well as their errors. These results are listed in Table ~\ref{table2}. An example of the continuum fit and the residual is shown in Fig.~\ref{fig1}.

\subsection{Measuring the variable regions in \civ BAL troughs}
In the \civ BAL tough for each QSO, it is possible to have some variable regions. To identify the variable regions in the \civ BAL-trough, we compare two-epoch spectra for BAL QSOs, measuring the flux deviation between two observations at each wavelength by the following equation \citep{Filiz13}:
\begin{equation}
N_{\sigma}(\lambda)=\frac{f_{1}-f_{2}}{\sqrt{\sigma_{1}^{2}+\sigma_{2}^{2}}}
\label{eq4}
\end{equation}
where $f_{1}$ and $f_{2}$ are the normalized flux based on the fitting power-law continuum and $\sigma_{1}$ and $\sigma_{2}$ are the normalized flux error at wavelength $\lambda$. Both $\sigma_{1}$ and $\sigma_{2}$ include observational flux errors and uncertainties on the continuum model. Similar to \cite{Gibson08}, variable regions of BAL troughs are identified to be where an absorption feature is detected with $|N_{\sigma}| \geq 1$ for at least five consecutive data points (4 \AA\ wide). This requirement allows detection of variable regions wider than 774 \kms. It is slightly smaller than that by \cite{Gibson08}, and larger than that by \cite{Filiz13}. For the number of data points to be lager than 5, the significance of variations would be $>99.9\%$. 73 variable regions in the \civ BAL troughs are identified from two-epoch different spectra in 43 BAL QSOs. For the sample of 188 BAL QSOs, there are about 23\% (43/188) BAL QSOs showing variable regions from two-epoch spectra. Fig.~\ref{fig3} gives an example of identified two variable regions in \civ BAL trough.

For each spectrum, we calculate EW and its error for each \civ BAL variable region (see above Eq.~\ref{eq2}, Eq.~\ref{eq3}). Then, we derive the EW variance for each \civ BAL variable region for each QSO. The sum of the EW variance for all variable regions in a QSO is adopted as its total EW variance, $\Delta EW$. The error of $\Delta EW$ is calculated based on the error propagation from errors of EW for all variable regions. For each variable region in \civ BAL troughs, the velocity width and the center outflow velocity are also calculated, as well as the maximum outflow velocity $V_{max}$ of variable regions. The velocity width is calculated from the left and right boundary of the variable region. In Table ~\ref{table3}, we list the total $\Delta EW$, the left and right boundary of the variable region for these 43 BAL QSOs.

\section{Results and discussion}

\subsection{Features of the variable regions: velocity width, outflow velocity, variation proportion}
\begin{figure}
\begin{center}

\includegraphics[height=8cm,angle=-90]{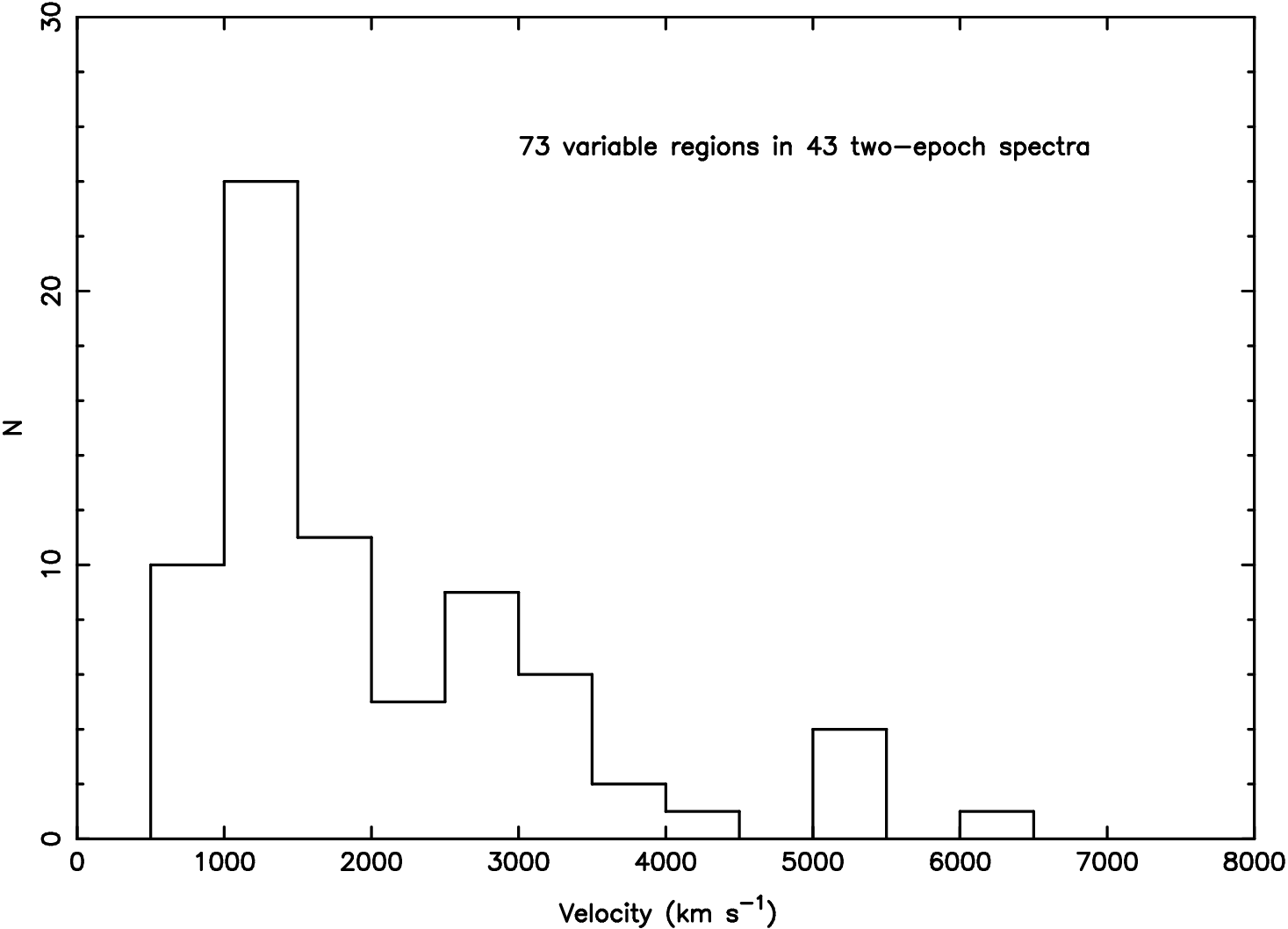}
\includegraphics[height=8cm,angle=-90]{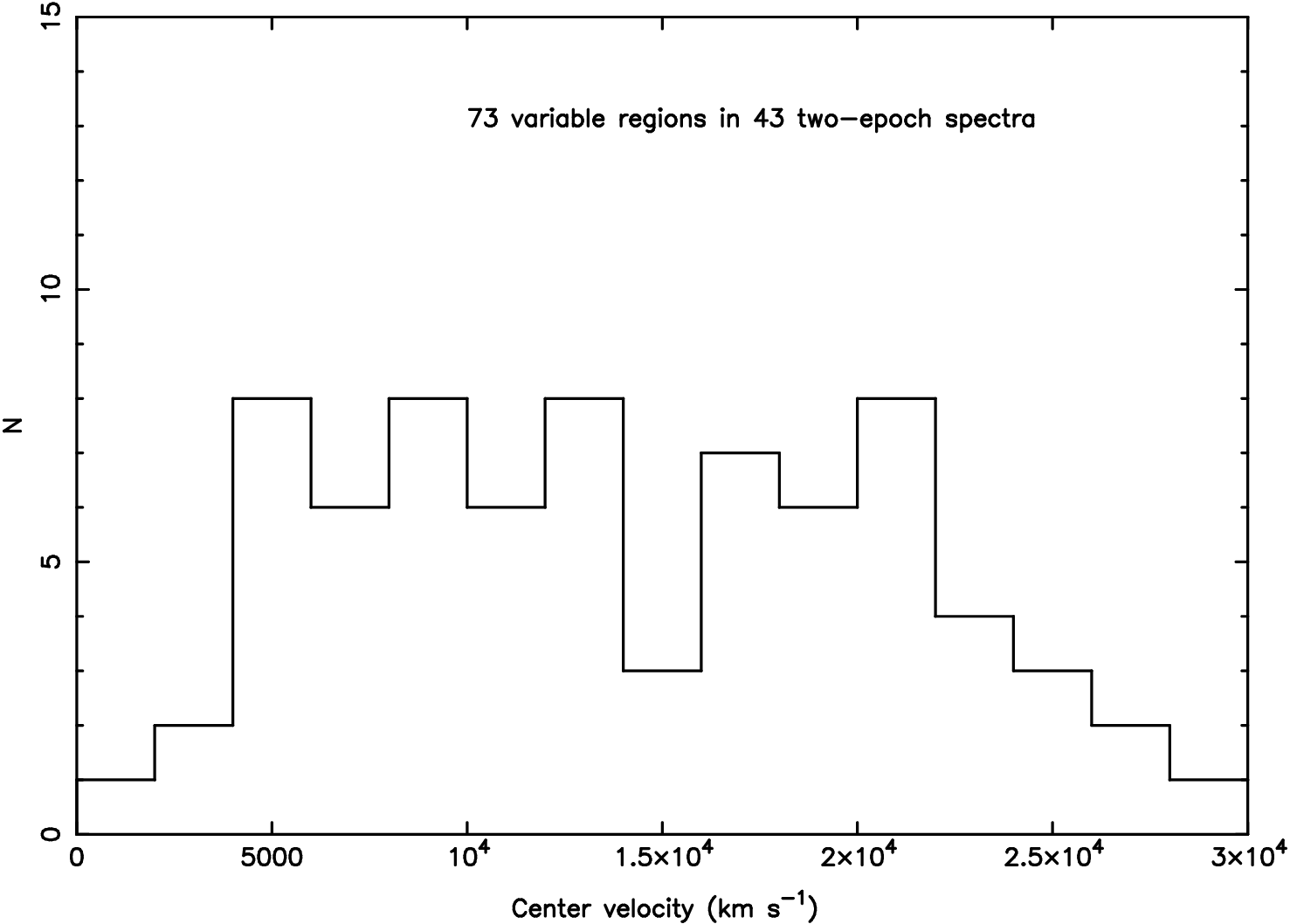}
\caption{Histograms for the 73 identified variable \civ BAL regions. Top: the distribution of the velocity width of the variable regions. Bottom: the distribution of the center outflow velocity of the variable regions. }
\label{fig4}
\end{center}
\end{figure}

With the above criterion of at least five consecutive data points larger than $1\sigma$, 73 variable regions in the \civ BAL troughs are identified from 43 two-epoch different spectra. Fig.~\ref{fig4} shows histograms of their velocity width and their center outflow velocity (Table ~\ref{table3}). BAL-troughs variation tends to occur on small velocity width. Even the largest variation widths (6500 \kms) are narrow compared to BAL-trough outflow velocity (bottom panel of Fig.~\ref{fig4}). The number of the \civ BAL variable regions decreases with the increase of the velocity width of the variable regions. These results are consistent with that by \cite{Gibson08}.

From the bottom panel of Fig.~\ref{fig4}, variable regions are found across a wide range of central outflow velocities, and the number of variable regions appears to peak in the range between 5000 and 21000 \kms. Considering $\rm BI_0$ used by \cite{Shen11}, $0~\kms$ is adopted as a minimum detection limit for BAL-trough. We do not consider contaminations from \civ 1549, \siiv 1399 emission lines at small outflow velocities and large outflow velocities, respectively. It could explain decrease of number of variable regions at small outflow velocities and at large outflow velocities in the bottom panel in Fig.~\ref{fig4}.

\begin{figure}
\begin{center}
\includegraphics[height=8cm,angle=-90]{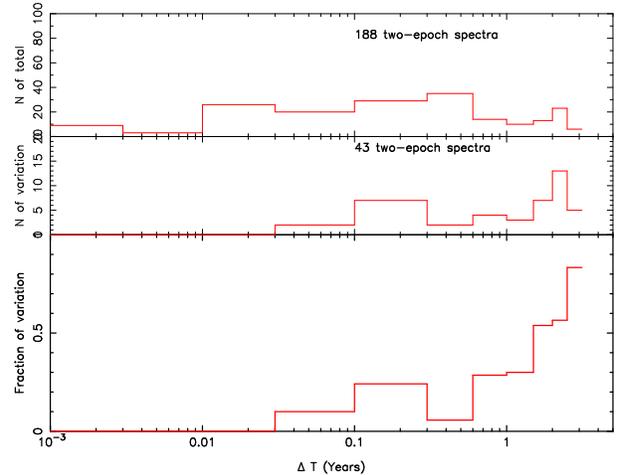}
\caption{Top: distribution of the time-interval between two epochs for all 188 BAL QSOs. Middle: distribution of the time-interval between two epochs for 43 \civ BAL QSOs with variable regions. Bottom: the number ratio of the middle panel to the top panel. The time-interval is in the rest frame of BAL QSOs.}
\label{fig5}
\end{center}
\end{figure}

In Fig.~\ref{fig5}, we give the distribution of the time-interval between epochs for total 188 BAL QSOs (top panel), as well as the time-interval histogram for 43 BAL QSOs showing variable regions (middle panel). It covers a time-interval of about 0.001 -- 3 years in the rest frame. The bottom panel in Fig.~\ref{fig5} is the number ratio of the middle panel to the top panel, i.e., the proportion of BAL QSOs showing variable regions versus the time-interval. It is clear that the proportion of BAL QSOs showing variable regions increases with the time-interval, rising to 50\% when the time-interval is longer than about 1 year in the rest frame. This result gives a good reason to consider only the longest time-interval spectral pairs for each QSO. Larger proportion of BAL trough showing variable region across longer time-interval is consistent with the result by \cite{Capellupo13}.

\subsection{The variability of spectral index $\Delta \alpha$ for BAL QSOs with variable regions}

\begin{table*}
\centering
\caption{Summary of the Spearman correlation
coefficients: For total sample of 188 BAL QSOs, $\Delta EW$ is the BAL-trough EW variation from two-epoch spectra, and there are 116 BAL QSOs with available \mgii-based \mbh. For 43 BAL QSOs with variable regions, $\Delta EW$ is the EW variation of the \civ BAL-trough variable regions, and there are 28 BAL QSOs with available \mgii-based \mbh. $\Delta L_{1500}$ is the variation of the continuum at 1500 \AA. $\Delta \alpha$ is the variation of the spectral index. $M_{\rm BH}$ is the \mgii-based SMBH masses from a single spectrum. $L_{bol}/L_{Edd}$ is the ratio of the bolometric luminosity to the Eddington luminosity. $V_{max}$ is the maximum velocity of the variable region. The value in brackets is the probability of the null hypothesis. }
\label{table1}
\begin{tabular}{lccccccccccccccccccc}
\hline
& $\Delta L_{1500}-\Delta \alpha$ & $\Delta EW -\Delta \alpha$ &$\Delta EW - \Delta L_{1500}$ & $\Delta EW - M_{BH}$ & $V_{max} - L_{bol}/L_{Edd}$ \\
\hline
Total sample & $-0.36(5.5E-07)$ &  $-0.008(0.91)$     &-0.17 (0.02)   & -0.15(0.1) & - \\
Subsample    & $-0.56(8.3E-05)$ &  $0.30(0.05)$      &-0.44 (0.003)  & -0.54(0.003) & 0.53 (0.004) \\
\hline
\end{tabular}
\end{table*}

\begin{figure}
\begin{center}
\includegraphics[height=8cm,angle=-90]{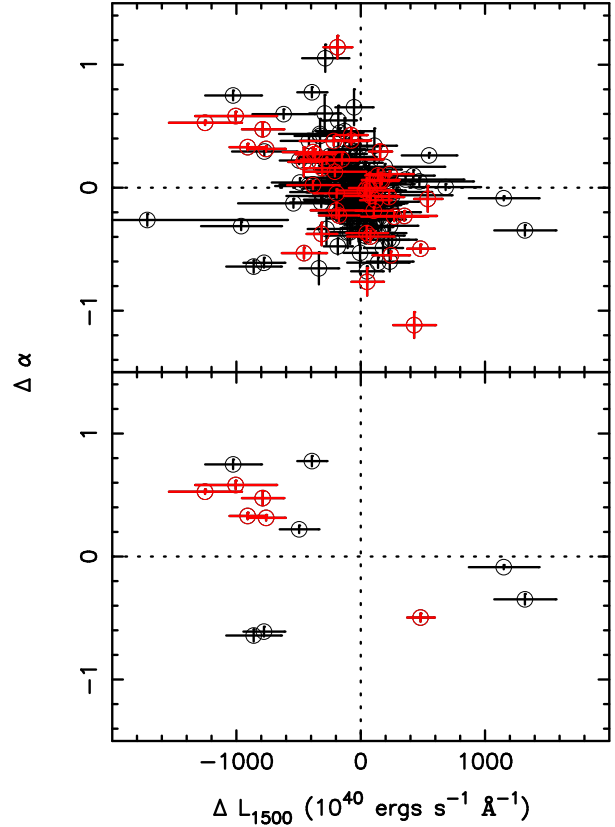}
\caption{$\Delta \alpha$ versus $\Delta L_{1500}$ for 188 BAL QSOs (top), and for 13 BAL QSOs with both  $\Delta L_{1500}$ and $\Delta \alpha$ more than 3$\sigma$ (bottom). The proportion of the points in the second and fourth quadrants (i.e., bluer during brighter phases) rises from about 56.9\% (top) to 84.6\% (bottom). The red circles denote 43 BAL QSOs with variable regions (top), and 6 QSOs with both  $\Delta L_{1500}$ and $\Delta \alpha$ more than 3$\sigma$ (bottom). The proportion of the points in the second and fourth quadrants is 76.7\%. For 6 QSOs with both $\Delta L_{1500}$ and $\Delta \alpha$ more than 3$\sigma$, the proportion of the points in the second and fourth quadrants rises to 100\%.}
\label{fig6}
\end{center}
\end{figure}

It has been shown that QSOs spectra are bluer during their brighter phases. It is usually explained by the variation of the accretion disk. Fig.~\ref{fig6} shows $\Delta \alpha$ versus the continuum variation at 1500\AA\ ($\Delta L_{1500}$) for total 188 BAL QSOs (the top panel). There is a significant weak correlation between them. The Spearman coefficient $R$ is $-0.36$, with the the probability of the null hypothesis of $P_{\rm null} = 5.5\times 10^{-7}$ (see Table ~\ref{table1}). In the second and fourth quadrants in the top panel in Fig.~\ref{fig6}, BAL QSOs show bluer during their brighter phase. There are about 56.9\% date points showing bluer during their brighter phase. It is consistent with the result for non-BAL QSOs by \cite{Bian12a}. Considering 11 points with both $\Delta L_{1500}$ and $\Delta \alpha$ more than 3$\sigma$, the proportion rises to 84.6\% (11/13), more QSOs appear bluer during their brighter phases compared to non-BAL QSOs (the bottom panel in Fig. ~\ref{fig6}).

For the subsample of BAL QSOs with variable regions in \civ BAL troughs, there is a significant medium strong correlation between the $\Delta L_{1500}$ and $\Delta \alpha$. The Spearman coefficient $R$ is $-0.56$, $P_{\rm null} = 8.3\times 10^{-5}$ (see Table 1). There are about 76.7\% BAL QSOs showing bluer during their brighter phases for 43 BAL QSOs (red circles in the top panel in Fig. ~\ref{fig6}). Considering 6 BAL QSOs with both $\Delta L_{1500}$ and $\Delta \alpha$ more than 3$\sigma$, the proportion of the points in the second and fourth quadrants rises to 100\% (6/6) (red circles in bottom panel in Fig. ~\ref{fig6}).

The proportion showing bluer during the brighter phases rises from about 56.9\% for total 188 BAL QSOs to about 76.7\% for 43 BAL QSOs with variable regions. During the brighter phases, the accretion disk becomes hotter and its emission peak would move to shorter wavelengths. It may lead to the trend of bluer spectra during brighter phases. Therefore, this larger proportion implies that the origin of variable BAL-trough regions is related to the central accretion process.




\subsection{The EW variability $\Delta EW$ for the variable region: the relation with $\Delta \alpha$, $\Delta L_{1500}$}

\begin{figure}
\begin{center}
\includegraphics[height=7cm,angle=-90]{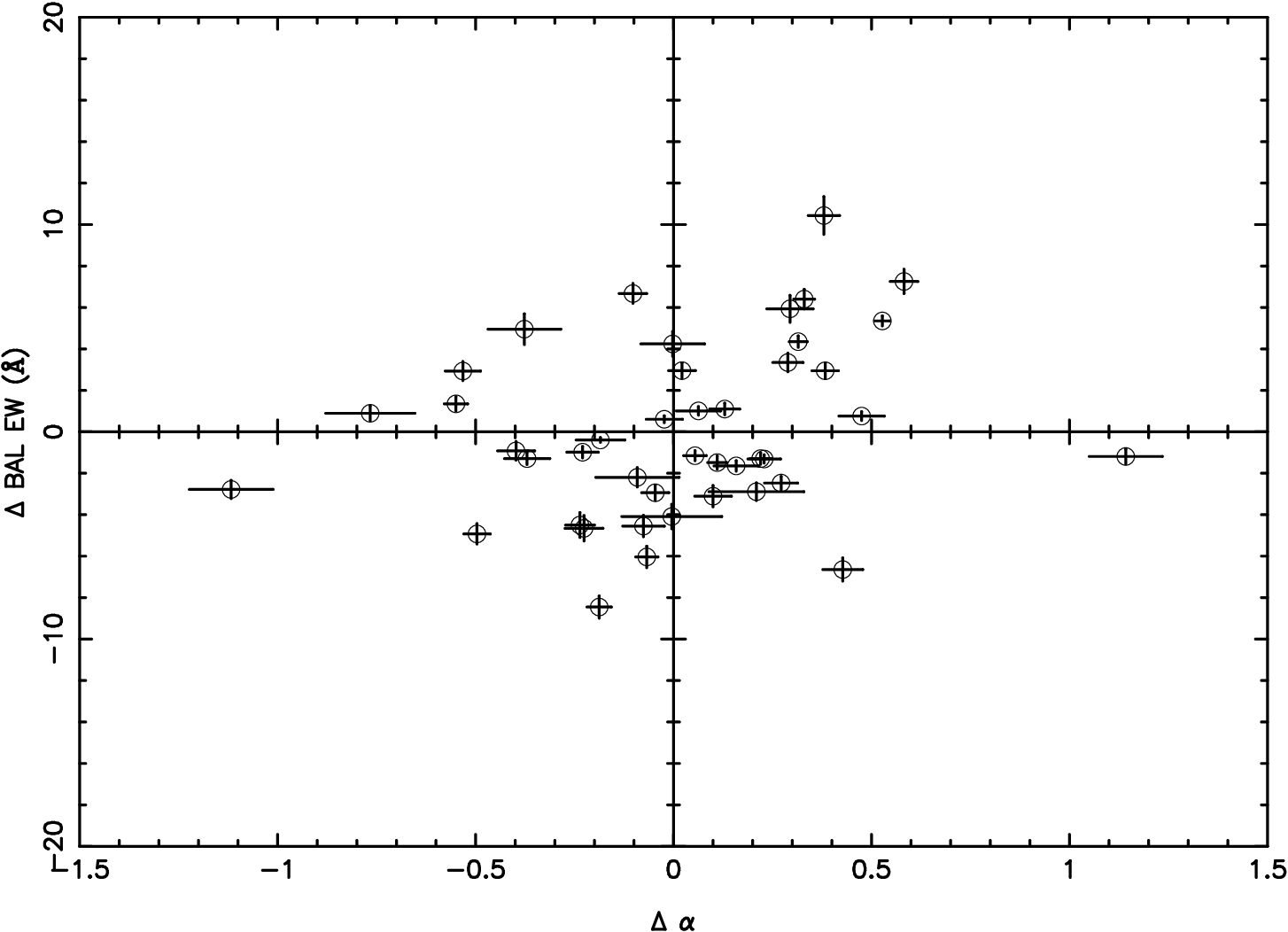}
\includegraphics[height=7cm,angle=-90]{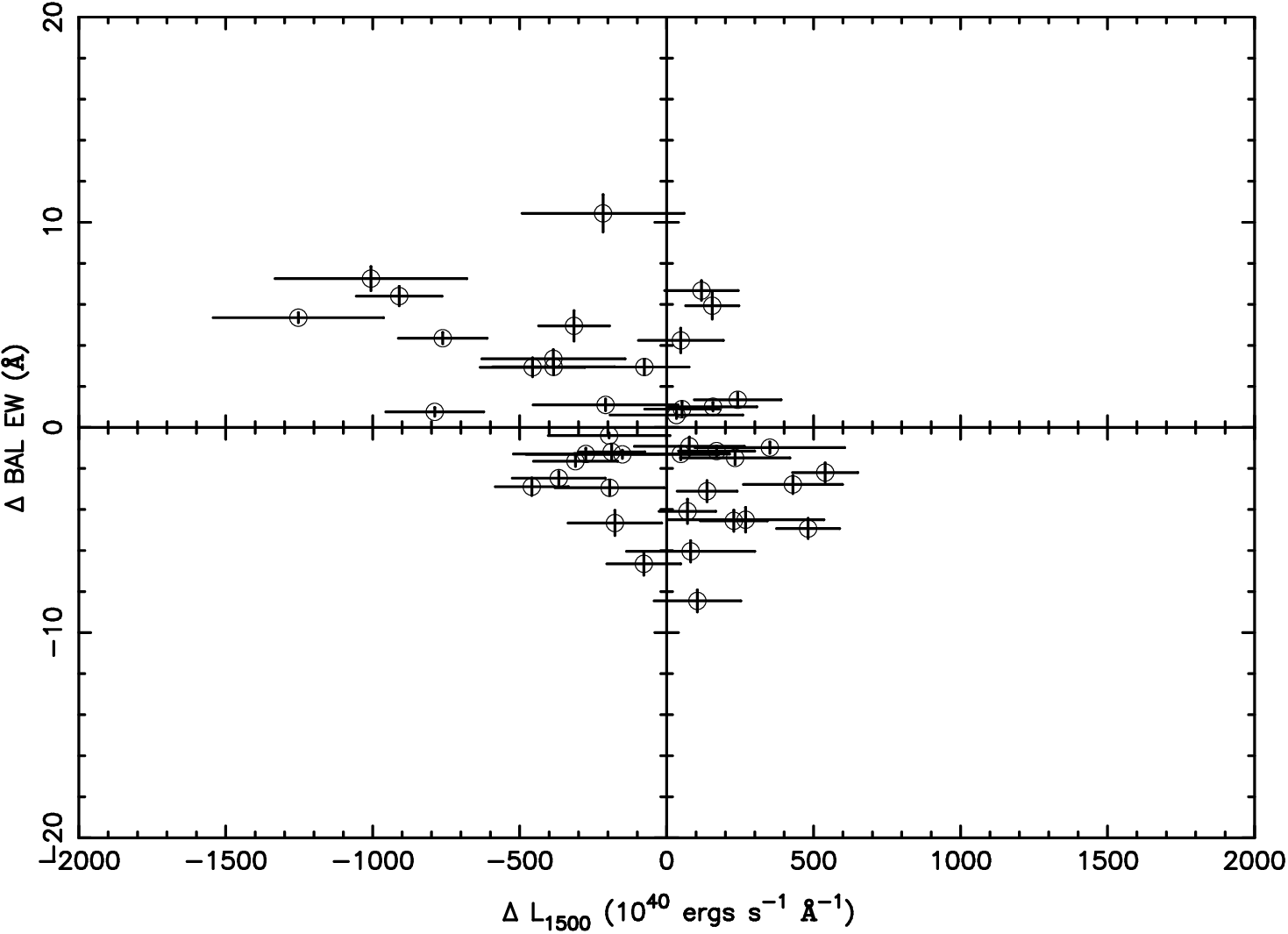}
\caption{Top: $\Delta EW $ versus $\Delta \alpha$ for 43 epochs in BAL QSOs with variable regions. The Spearman correlation coefficient $R$ is $0.3$ with $P_{\rm null} = 0.05$. Bottom: $\Delta EW $ versus $\Delta L_{5100}$ for 43 epochs in BAL QSOs with variable regions, $R=-0.44, P_{\rm null} = 0.003$. }
\label{fig7}
\end{center}
\end{figure}

For the subsample of 43 BAL QSOs with variable regions in \civ BAL troughs, the relation between $\Delta EW$ and $\Delta \alpha$, $\Delta L_{1500}$ are used to investigate the origin of BAL variability (see Table ~\ref{table2}, Table ~\ref{table3}). Fig. ~\ref{fig7} shows $\Delta EW$ versus $\Delta \alpha$ (top panel). It is possible that there is a weak correlation between $\Delta EW$ and $\Delta \alpha$, according to a Spearman rank correlation test ($R=0.3, P_{\rm null} = 0.05$; Table ~\ref{table1}). There are 60.5\% of QSOs located in the first and third quadrants where the BAL-troughs become stronger when BAL QSOs become redder. This is different from our previous study for a single BAL QSOs with 18 epochs \citep{He14}. \cite{He14} found a strong correlation between EW variation for \civ BAL-trough and $\Delta \alpha$ (R=0.77), suggesting that dust is intrinsic to outflows. For the subsample of BAL QSOs with variable regions, we don't confirm that strong correlation from the two-epoch variation. It is possibly due to the additional parameters to blur this correlation. For the total sample of 188 BAL QSOs, there is no correlation between $\Delta EW$ and $\Delta \alpha$, where $\Delta EW$ is the EW variation for \civ BAL-trough. A Spearman rank correlation test gives $R=-0.008, P_{\rm null} = 0.91$ (Table ~\ref{table1}).

The bottom panel in Fig.~\ref{fig7} shows $\Delta EW$ versus $\Delta L_{1500}$ for 43 BAL QSOs with variable regions. It is possible that there is a medium strong correlation between them, $R=-0.44, P_{\rm null} = 0.003$ (Table ~\ref{table1}). It is found that 60.5\% QSOs located in the second and fourth quadrants, where BAL troughs become weaker when BAL QSOs become brighter. For the total sample of 188 BAL QSOs, $\Delta EW$ is the EW variation for \civ BAL-trough, and $R=-0.17, P_{\rm null} = 0.02$ (Table ~\ref{table1}). Therefore, we don't find a significant correlation between them. It is consistent with the study of \cite{Filiz13}, who found no significant evidence for EW variability of the \civ BAL trough driven by QSO bolometric luminosity \citep[also see][]{Gibson08}. With respect to total sample of 188 BAL QSOs, the subsample of 43 BAL QSOs with variable regions has a stronger correlation between them. Considering the probability of the null hypothesis ($P_{\rm null} = 0.003$) for the subsample of 43 BAL QSOs with variable regions, this correlation is suggestive and  requires further investigations by larger sample.



\subsection{The relation of $\Delta EW$ with the central accretion properties, \mbh, \leddR}
The relation between the BAL properties/variability and the accretion process in BAL QSOs has been discussed by many authors \citep[e.g.][]{Laor02, Ganguly07, Baskin13, Filiz13, He14}. There are two fundamental parameters related to the accretion process, the SMBH mass (\mbh) and the Eddington ratio (\leddR). Considering that the \civ-based SMBH mass is biased to the possible non-virialized component in \civ emission lines \citep[e.g.][]{Rafiee11, Shen11, Bian12b}, we used \mgii-based \mbh to investigate the relation of $\Delta EW$ with the central accretion properties, such as \mbh, and \leddR. Fitting \mgii\ 2800 emission lines, \cite{Shen11} gave the \mgii-based \mbh for QSOs with $ 0.35 \leq z \leq 2.25$. For the subsample of BAL QSOs with variable regions in \civ BAL toughs, there are 28 BAL QSOs with measured \mgii-based \mbh by \cite{Shen11}. The bolometric luminosity (\lb) is calculated from $L_{3000}$ ($0.7\leq z\leq 1.9$), $L_{1350}$ ($ z\geq 1.9$). With the bolometric corrections (BC) from the luminosity at 3000 \AA\, and 1350 \AA\ as $BC_{3000}=5.15$, $BC_{1350}=3.81$, we calculate \leddR. These results are listed in Table ~\ref{table3}.

\begin{figure}
\begin{center}
\includegraphics[height=8cm,angle=-90]{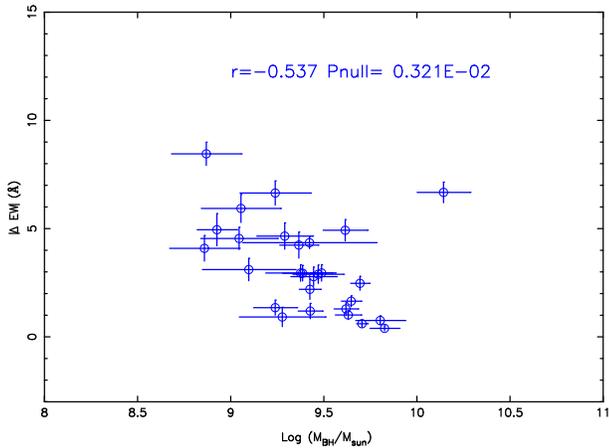}
\caption{ $|\Delta EW|$ versus the \mgii-based \mbh for 28 BAL QSOs with variable regions. The Spearman correlation coefficient $R$ is $-0.54$ with $P_{null} = 0.003$.
}
\label{fig8}
\end{center}
\end{figure}

\begin{figure}
\begin{center}
\includegraphics[height=8cm,angle=-90]{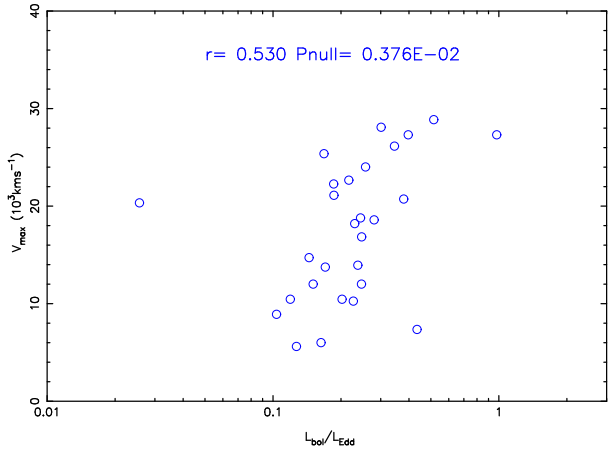}
\caption{The maximum outflow velocity of variable regions ($V_{max}$) versus the Eddington ratio \leddR for 28 BAL QSOs with variable regions. The Spearman correlation coefficient $R$ is $0.53$ with $P_{null} = 0.004$.}
\label{fig9}
\end{center}
\end{figure}

In Fig.~\ref{fig8}, we show the relation between $|\Delta EW|$ and \mgii-based \mbh. It is possible that there is a negative medium strong correlation between the  $|\Delta EW|$ and \mbh ($R=-0.54, P_{null}=0.003$, Table ~\ref{table1}), showing smaller variation of BAL-trough for BAL QSOs with larger SMBHs masses. In Fig. ~\ref{fig9}, we shows the relation between the maximum outflow velocity of variable regions ($V_{max}$) and the Eddington ratio \leddR. It is possible that there is a medium strong correlation between the $V_{max}$ and \leddR ($R=0.53, P_{null}=0.004$, Table ~\ref{table1}), showing larger outflow velocity for BAL QSOs with larger Eddington ratio. Considering the test confidence less than 99.9\% for the subsample of 43 BAL QSOs, these correlations are suggestive. For the total 188 BAL QSOs, there are 116 BAL QSOs with available \mgii-based \mbh, and there is no significant correlation between $|\Delta EW|$ and \mbh ($R=-0.15, P_{null}=0.1$,  Table ~\ref{table1}).

In Figs.~\ref{fig8},~\ref{fig9}, there is an outlier, SDSS J020006.31-003709.7 ($z=2.141$), with largest \mbh and smallest \leddR in Table ~\ref{table3}. Excluding this QSO, these correlations would be stronger and significant. For the relation between $|\Delta EW|$ and \mbh, $R=-0.7, P_{null}=0.00005$. For the relation between $V_{max}$ and \leddR, $R=0.58, P_{null}=0.0014$. These results imply the connection between the BAL-trough variation and the central accretion process. Using \civ-based SMBH mass for BAL QSOs ($z>2$), \cite{Filiz13} suggested possible correlations between $|\Delta EW|$ and luminosity/Eddington ratio for \civ troughs on moderate time-scales (1-2.5 yrs). They did not find significant correlation between $|\Delta EW|$ and \mbh. These correlations require further investigations in larger samples of BAL QSOs with no bias \mbh estimation such as from \hb, \mgii\ emission lines.

We don't find a significant correlation between $V_{max}$ and $\alpha$, \lb. We don't confirm the result by \cite{Laor02}, who found that $V_{max}$  increase with \lb. We find that \leddR instead of \lb is a driver of  $V_{max}$ . In the future, we can use the BOSS spectra to estimate the \mbh because of their larger wavelength coverage in BOSS spectra than in SDSS spectra.


\section{SUMMARY}
The variability of broad absorption lines is investigated for a sample of 188 BAL QSOs ($z > 1.7$) with at least two-epoch observations from the SDSS DR7. Considering only the longest time-scale between epochs for each QSO, 73 variable regions in the \civ BAL troughs are detected for 43 BAL QSOs. The main conclusions can be summarized as follows.

(1) Considering the flux deviation $|N_{\sigma}| \geq 1$ for at least five consecutive data points (4 \AA\ wide) in the two-epoch difference spectrum, 73 \civ BAL variable regions in 43 BAL QSOs are identified from 188 BAL QSOs, i.e., about 23\% (43/188) BAL QSOs showing variable regions from two-epoch spectra. It is found that
the proportion of BAL QSOs showing variable regions increases with the time-interval longer than about 1 year in the rest frame. BAL variation tends to occur on small velocity width. Even the largest variation widths (6500 \kms) are narrow compared to BAL-trough outflow velocity. Variable regions are found across a wide range of outflow velocities, and the number of variable regions appears to peak in the range between $\sim$ 5000 and $\sim$ 21000 \kms.

(2) With two-epoch variation, it is found that there is a weak correlation between $\Delta L_{1500}$ and $\Delta \alpha$ for total 188 two-epoch spectra, and about half BAL QSOs appear redder during their brighter phases. It is consistent with the result for non-BAL QSOs by \citep{Bian12a}. For a subsample of BAL QSOs with variable  regions in BAL toughs, their correlation becomes stronger, about 76.7\% BAL QSOs appear bluer during their brighter phases. This larger proportion implies that the origin of variable regions in BAL-trough is related to the central accretion process.

(3) For the subsample of 43 BAL QSOs with variable regions, it is possible that there is a weak correlation between $\Delta EW$ and $\Delta \alpha$. It is different from our previous study for a single BAL QSOs by \citep{He14}. It is possibly due to the additional parameters to blur this correlation. It is possible that there is a medium strong correlation between $\Delta EW$ and $\Delta L_{5100}$. Considering the test confidence is 99.7\%, this latter correlation is suggestive and requires further investigations by larger sample.

(4) For the subsample of 43 BAL QSOs with variable regions, there are 28 BAL QSOs with available \mgii-based \mbh. It is possible that there is a negative medium strong correlation between the $|\Delta EW|$ and \mbh, showing smaller variation of BAL-trough for BAL QSOs with larger SMBHs masses. It is possible that there is a medium strong correlation between $V_{max}$ of variable regions and \leddR, showing larger outflow velocity for BAL QSOs with larger Eddington ratio. These results imply the connection between the BAL-trough variation and the central accretion process. These correlations require further investigations in larger samples of BAL QSOs with no bias \mbh estimation such as from \hb, \mgii\ emission lines.

\section{ACKNOWLEDGMENTS}
We are very grateful to the anonymous referee for her/his instructive comments which significantly improved the content of the paper. This work has been supported by the National Science Foundations of China (Nos. 11373024, 11173016 and 11233003).

\setcounter{table}{1}
\begin{center}
\begin{table*}
\centering \caption{The properties of 188 BAL QSOs. Col. (1) is the No. of BAL QSOs. Col. (2) is No. of the epoch. Col. (3) is the SDSS name. Col. (4) is the redshift. Col. (5) is Plate-MJD-Fiber of the observation. Col. (6) is luminosity in units of $10^{40} \ergs \AA^{-1}$. Col. (7) is the UV spectral index. Col. (8) is the total \civ BAL EW in units of \AA.}
\label{table2}
\begin{tabular}{lccccccccccccccccccc}

\hline \hline
  No.  & epoch &name  & z& Plate-MJD-Fiber &$L_{1500}$ & $\alpha$ & BAL EW \\
  (1) & (2) & (3) & (4) & (5) & (6) & (7) & (8)\\
\hline
001&01&J001130.55+005550.7 & 2.309&0389-51795-0339&$ 1860.73\pm 220.18$&$-1.18\pm 0.04$&$ 12.03\pm 1.21$\\
001&02&J001130.55+005550.7 & 2.309&0686-52519-0603&$ 1475.93\pm 154.57$&$-1.15\pm 0.03$&$ 10.18\pm 1.10$\\
002&01&J001438.28-010750.1 & 1.806&0389-51795-0211&$  770.10\pm 148.78$&$-0.88\pm 0.13$&$ 24.11\pm 2.07$\\
002&02&J001438.28-010750.1 & 1.806&0687-52518-0249&$  615.19\pm 128.45$&$-0.81\pm 0.13$&$ 21.59\pm 2.29$\\
003&01&J002127.88+010420.1 & 1.820&0390-51816-0445&$ 1604.75\pm 143.90$&$-1.75\pm 0.08$&$ 14.41\pm 0.98$\\
003&02&J002127.88+010420.1 & 1.820&0688-52203-0374&$ 1349.17\pm 129.10$&$-1.70\pm 0.08$&$ 14.40\pm 1.04$\\
004&01&J002146.71-004847.9 & 2.503&0390-51816-0161&$ 1824.74\pm 183.96$&$-1.22\pm 0.04$&$  5.35\pm 1.10$\\
004&02&J002146.71-004847.9 & 2.503&0390-51900-0180&$ 2074.63\pm 134.06$&$-1.42\pm 0.02$&$  4.82\pm 0.69$\\
005&01&J002710.06-094435.3 & 2.070&0653-52145-0556&$ 1869.96\pm 126.53$&$-0.76\pm 0.03$&$ 16.80\pm 0.75$\\
005&02&J002710.06-094435.3 & 2.070&3105-54825-0310&$ 1974.66\pm  75.48$&$-0.95\pm 0.01$&$ 10.82\pm 0.41$\\
006&01&J003312.25+155442.4 & 1.937&0417-51821-0576&$  770.90\pm  88.62$&$-1.20\pm 0.12$&$ 69.25\pm 1.26$\\
006&02&J003312.25+155442.4 & 1.937&3133-54789-0379&$  841.70\pm  38.27$&$-1.21\pm 0.05$&$ 62.26\pm 0.51$\\
007&01&J004118.59+001742.4 & 1.765&0392-51793-0631&$  723.13\pm 113.58$&$-1.05\pm 0.10$&$ 16.72\pm 1.55$\\
007&02&J004118.59+001742.4 & 1.765&0690-52261-0553&$  712.34\pm  91.32$&$-1.38\pm 0.10$&$ 13.40\pm 1.41$\\
008&01&J004323.43-001552.4 & 2.798&0393-51794-0181&$ 3052.29\pm 367.63$&$-0.86\pm 0.02$&$  9.68\pm 1.03$\\
008&02&J004323.43-001552.4 & 2.798&0690-52261-0028&$ 2511.14\pm 252.10$&$-0.99\pm 0.03$&$ 12.28\pm 1.17$\\
009&01&J004527.68+143816.1 & 1.992&0419-51812-0105&$ 3911.52\pm 164.76$&$-0.27\pm 0.03$&$ 27.37\pm 0.52$\\
009&02&J004527.68+143816.1 & 1.992&0419-51868-0106&$ 3944.77\pm 154.36$&$-0.30\pm 0.03$&$ 28.71\pm 0.48$\\
010&01&J004613.54+010425.7 & 2.152&0393-51794-0572&$ 2955.44\pm 172.22$&$-1.66\pm 0.02$&$ 39.71\pm 0.54$\\
010&02&J004613.54+010425.7 & 2.152&0691-52199-0460&$ 2858.19\pm 131.32$&$-1.67\pm 0.02$&$ 37.86\pm 0.42$\\
011&01&J004732.73+002111.3 & 2.878&0393-51794-0588&$ 2669.00\pm 276.06$&$-1.91\pm 0.03$&$  8.81\pm 1.12$\\
011&02&J004732.73+002111.3 & 2.878&0691-52199-0559&$ 2749.06\pm 207.40$&$-1.89\pm 0.02$&$  8.81\pm 0.83$\\
012&01&J004806.05-010321.6 & 2.528&0394-51812-0299&$ 3502.07\pm 173.05$&$-1.73\pm 0.02$&$  4.79\pm 0.55$\\
012&02&J004806.05-010321.6 & 2.528&3111-54800-0281&$ 4051.75\pm 136.42$&$-1.47\pm 0.01$&$  3.70\pm 0.37$\\
013&01&J005419.99+002727.9 & 2.490&0394-51812-0511&$ 2561.01\pm 152.81$&$-1.13\pm 0.02$&$ 16.86\pm 0.62$\\
013&02&J005419.99+002727.9 & 2.490&3111-54800-0509&$ 2793.79\pm 106.36$&$-1.02\pm 0.01$&$ 13.95\pm 0.39$\\
014&01&J010859.53-105757.8 & 1.807&0659-52199-0099&$ 1260.76\pm 105.13$&$-1.18\pm 0.07$&$ 48.50\pm 1.02$\\
014&02&J010859.53-105757.8 & 1.807&3109-54833-0003&$  945.67\pm  58.56$&$-1.56\pm 0.06$&$ 52.70\pm 0.72$\\
015&01&J010920.96-100127.0 & 2.449&2864-54467-0320&$ 1277.23\pm  72.02$&$-1.68\pm 0.02$&$  9.86\pm 0.63$\\
015&02&J010920.96-100127.0 & 2.449&3109-54833-0080&$  884.09\pm  93.49$&$-0.91\pm 0.04$&$  9.54\pm 1.17$\\
016&01&J012603.62-100114.8 & 2.302&0661-52163-0114&$ 1512.19\pm 160.55$&$-1.95\pm 0.03$&$ 22.54\pm 1.06$\\
016&02&J012603.62-100114.8 & 2.302&2878-54465-0274&$ 1055.95\pm  77.02$&$-2.48\pm 0.03$&$ 31.04\pm 0.79$\\
017&01&J013012.36+153158.0 & 2.352&0425-51898-0445&$ 1778.04\pm 174.69$&$ 0.07\pm 0.03$&$ 20.49\pm 1.06$\\
017&02&J013012.36+153158.0 & 2.352&0425-51884-0446&$ 1769.05\pm 189.22$&$-0.18\pm 0.04$&$ 23.50\pm 1.15$\\
018&01&J013038.79+391818.1 & 2.520&2062-53381-0623&$ 2139.81\pm 119.40$&$-1.27\pm 0.02$&$ 13.34\pm 0.60$\\
018&02&J013038.79+391818.1 & 2.520&2063-53359-0024&$ 2024.22\pm 121.10$&$-1.39\pm 0.02$&$ 12.69\pm 0.64$\\
019&01&J013625.65-103346.2 & 2.009&0662-52147-0002&$  799.65\pm 121.52$&$-0.14\pm 0.08$&$ 42.02\pm 1.68$\\
019&02&J013625.65-103346.2 & 2.009&1915-53612-0133&$  650.34\pm  75.17$&$-0.27\pm 0.05$&$ 42.34\pm 1.26$\\
020&01&J013656.31-004623.7 & 1.716&0400-51820-0003&$ 1338.49\pm 183.80$&$-1.10\pm 0.09$&$ 13.55\pm 1.52$\\
020&02&J013656.31-004623.7 & 1.716&0698-52203-0202&$ 1158.39\pm 124.19$&$-0.83\pm 0.06$&$ 19.78\pm 1.15$\\
021&01&J014948.74+141300.9 & 2.188&0429-51820-0584&$ 1745.76\pm 104.38$&$-1.43\pm 0.02$&$  7.85\pm 0.62$\\
021&02&J014948.74+141300.9 & 2.188&1899-53262-0548&$ 1740.28\pm  82.23$&$-1.53\pm 0.02$&$  7.84\pm 0.51$\\
022&01&J015048.83+004126.2 & 3.702&0402-51793-0505&$ 5558.73\pm 320.86$&$-1.30\pm 0.04$&$ 10.18\pm 0.63$\\
022&02&J015048.83+004126.2 & 3.702&0699-52202-0629&$ 5319.60\pm 315.36$&$-1.29\pm 0.05$&$ 12.07\pm 0.68$\\
023&01&J020006.31-003709.7 & 2.141&0403-51871-0070&$ 1846.41\pm 112.19$&$-0.41\pm 0.03$&$ 40.36\pm 0.73$\\
023&02&J020006.31-003709.7 & 2.141&2866-54478-0197&$ 1964.77\pm  54.49$&$-0.51\pm 0.02$&$ 48.21\pm 0.39$\\
024&01&J021818.14-092153.5 & 1.880&0668-52162-0218&$ 2259.30\pm 149.56$&$-1.62\pm 0.05$&$ 16.67\pm 0.73$\\
024&02&J021818.14-092153.5 & 1.880&3122-54821-0201&$ 1470.26\pm  73.36$&$-1.14\pm 0.03$&$ 19.90\pm 0.51$\\
025&01&J022036.27-081242.9 & 2.004&0668-52162-0547&$ 2005.37\pm 129.03$&$-0.89\pm 0.03$&$ 20.99\pm 0.71$\\
025&02&J022036.27-081242.9 & 2.004&3122-54821-0482&$ 2247.01\pm  71.30$&$-1.44\pm 0.01$&$ 20.66\pm 0.36$\\
026&01&J022349.24+004727.8 & 2.335&0704-52205-0459&$  912.86\pm 100.24$&$-1.24\pm 0.09$&$ 13.93\pm 1.19$\\
026&02&J022349.24+004727.8 & 2.335&3127-54835-0338&$  964.99\pm  77.56$&$-2.01\pm 0.07$&$ 19.04\pm 0.82$\\
027&01&J022844.09+000217.0 & 2.705&0406-51817-0035&$ 3900.33\pm 187.14$&$-1.02\pm 0.02$&$ 18.87\pm 0.70$\\
027&02&J022844.09+000217.0 & 2.705&0406-52238-0039&$ 3854.36\pm 202.67$&$-0.94\pm 0.02$&$ 22.61\pm 0.73$\\
028&01&J023139.53+001758.4 & 2.375&0407-51820-0483&$ 1583.17\pm 118.98$&$-1.82\pm 0.03$&$  9.29\pm 0.87$\\
028&02&J023139.53+001758.4 & 2.375&0705-52200-0505&$ 1372.57\pm 107.43$&$-1.82\pm 0.03$&$ 12.73\pm 0.94$\\
029&01&J023252.80-001351.1 & 2.033&0705-52200-0063&$ 1231.68\pm 111.34$&$-1.95\pm 0.04$&$ 26.64\pm 1.08$\\
029&02&J023252.80-001351.1 & 2.033&3126-54804-0205&$ 1153.97\pm  57.90$&$-1.53\pm 0.02$&$ 17.54\pm 0.58$\\
030&01&J023820.90+001419.7 & 2.782&0706-52199-0472&$ 2195.28\pm 223.61$&$-1.94\pm 0.03$&$ 22.57\pm 1.28$\\
030&02&J023820.90+001419.7 & 2.782&3126-54804-0500&$ 1878.07\pm 115.63$&$-2.06\pm 0.02$&$ 18.17\pm 0.72$\\

\hline
\end{tabular}
\end{table*}
\end{center}

\setcounter{table}{1}
\begin{table*}
\centering \caption{-- continue}
\begin{tabular}{lccccccccccccccccccc}

\hline \hline
  No.  & epoch &name  & z& Plate-MJD-Fiber &$L_{1500}$ & $\alpha$ & BAL EW \\
  (1) & (2) & (3) & (4) & (5) & (6) & (7) & (8)\\
\hline
031&01&J023903.43-003850.8 & 3.075&0408-51821-0134&$ 2955.93\pm 219.75$&$-1.50\pm 0.03$&$  6.99\pm 0.98$\\
031&02&J023903.43-003850.8 & 3.075&0706-52199-0179&$ 2807.41\pm 207.90$&$-1.61\pm 0.03$&$  7.91\pm 0.92$\\
032&01&J024221.87+004912.6 & 2.069&0408-51821-0576&$ 2171.53\pm 137.44$&$-1.00\pm 0.02$&$ 19.70\pm 0.63$\\
032&02&J024221.87+004912.6 & 2.069&0706-52199-0617&$ 2028.23\pm 147.32$&$-0.98\pm 0.03$&$ 21.83\pm 0.71$\\
033&01&J024224.02+010452.5 & 2.433&0408-51821-0564&$ 1394.86\pm 152.27$&$-0.61\pm 0.04$&$ 12.18\pm 1.16$\\
033&02&J024224.02+010452.5 & 2.433&0706-52199-0606&$ 1065.54\pm 137.63$&$-0.41\pm 0.04$&$ 11.56\pm 1.37$\\
034&01&J024304.68+000005.4 & 1.995&0408-51821-0080&$ 1674.30\pm 135.82$&$-1.60\pm 0.03$&$ 13.76\pm 0.83$\\
034&02&J024304.68+000005.4 & 1.995&0706-52199-0080&$ 1464.74\pm 133.02$&$-1.78\pm 0.04$&$ 14.05\pm 1.00$\\
035&01&J024701.18+000330.2 & 2.152&0409-51871-0472&$  745.43\pm 118.16$&$ 0.11\pm 0.06$&$  8.28\pm 1.62$\\
035&02&J024701.18+000330.2 & 2.152&1664-52973-0474&$  659.55\pm  48.46$&$ 0.21\pm 0.03$&$  6.50\pm 0.79$\\
036&01&J025042.45+003536.7 & 2.385&0410-51816-0352&$ 1730.23\pm 157.63$&$-0.49\pm 0.03$&$ 40.90\pm 1.02$\\
036&02&J025042.45+003536.7 & 2.385&1664-52973-0548&$  951.55\pm  50.05$&$-1.10\pm 0.02$&$ 43.20\pm 0.58$\\
037&01&J025331.93+001624.6 & 1.821&0410-51816-0391&$ 1174.54\pm 155.96$&$-2.00\pm 0.10$&$  9.37\pm 1.44$\\
037&02&J025331.93+001624.6 & 1.821&0708-52175-0472&$ 1130.18\pm  95.74$&$-2.01\pm 0.07$&$  5.25\pm 0.95$\\
038&01&J031331.22-070422.8 & 2.786&0459-51924-0490&$ 2468.98\pm 220.86$&$-1.85\pm 0.03$&$ 41.52\pm 1.03$\\
038&02&J031331.22-070422.8 & 2.786&3185-54829-0623&$ 1848.56\pm 120.53$&$-1.25\pm 0.02$&$ 42.18\pm 0.73$\\
039&01&J031609.83+004043.1 & 2.897&0413-51821-0386&$ 3229.25\pm 401.19$&$-1.16\pm 0.03$&$  9.04\pm 1.19$\\
039&02&J031609.83+004043.1 & 2.897&3183-54833-0465&$ 3258.12\pm 164.31$&$-1.28\pm 0.01$&$  8.83\pm 0.46$\\
040&01&J031828.90-001523.1 & 1.985&0413-51821-0166&$ 3089.15\pm 166.78$&$-1.82\pm 0.02$&$  7.94\pm 0.59$\\
040&02&J031828.90-001523.1 & 1.985&0711-52202-0114&$ 2600.48\pm 154.31$&$-1.78\pm 0.03$&$ 10.64\pm 0.65$\\
041&01&J032118.22-010539.9 & 2.412&0413-51821-0055&$ 2625.51\pm 193.52$&$-0.12\pm 0.03$&$ 23.14\pm 0.85$\\
041&02&J032118.22-010539.9 & 2.412&0712-52199-0287&$ 1598.91\pm 115.59$&$ 0.63\pm 0.03$&$ 18.29\pm 0.87$\\
042&01&J032701.43-002207.1 & 2.319&0414-51869-0152&$ 1666.84\pm 151.69$&$-1.43\pm 0.03$&$ 12.03\pm 0.90$\\
042&02&J032701.43-002207.1 & 2.319&0712-52199-0159&$ 1727.30\pm 179.78$&$-1.60\pm 0.03$&$ 11.20\pm 1.00$\\
043&01&J033029.75-005918.1 & 2.747&0414-51869-0004&$ 1860.40\pm 228.82$&$-1.23\pm 0.04$&$  9.48\pm 1.55$\\
043&02&J033029.75-005918.1 & 2.747&0414-51901-0021&$ 1654.68\pm 195.10$&$-0.98\pm 0.04$&$ 20.70\pm 1.62$\\
044&01&J074123.74+190453.8 & 2.288&2074-53437-0325&$  866.49\pm  71.04$&$-1.08\pm 0.03$&$  7.66\pm 0.86$\\
044&02&J074123.74+190453.8 & 2.288&2915-54497-0338&$  852.48\pm  73.87$&$-1.08\pm 0.03$&$ 11.03\pm 0.90$\\
045&01&J074221.38+165740.3 & 2.538&2074-53437-0207&$ 1383.77\pm  85.90$&$-1.31\pm 0.02$&$  6.93\pm 0.66$\\
045&02&J074221.38+165740.3 & 2.538&2915-54497-0216&$ 1553.70\pm  96.80$&$-1.25\pm 0.02$&$  5.10\pm 0.68$\\
046&01&J074818.62+272304.7 & 2.344&2075-53737-0043&$  658.23\pm  76.70$&$-1.29\pm 0.04$&$  7.46\pm 1.24$\\
046&02&J074818.62+272304.7 & 2.344&2075-53730-0055&$  679.43\pm  58.28$&$-1.31\pm 0.03$&$  7.36\pm 0.94$\\
047&01&J075007.63+275708.0 & 2.364&1059-52618-0071&$ 2710.02\pm 128.52$&$-1.36\pm 0.02$&$ 23.93\pm 0.62$\\
047&02&J075007.63+275708.0 & 2.364&2075-53737-0550&$ 1799.98\pm  68.30$&$-1.03\pm 0.02$&$ 29.12\pm 0.50$\\
048&01&J080049.70+092830.7 & 2.182&2419-54139-0262&$ 1248.89\pm 130.64$&$-1.26\pm 0.04$&$  9.03\pm 1.15$\\
048&02&J080049.70+092830.7 & 2.182&2945-54505-0638&$ 1316.89\pm  72.55$&$-1.17\pm 0.02$&$ 11.08\pm 0.60$\\
049&01&J080455.90+231501.8 & 2.162&1265-52705-0187&$ 1467.36\pm 153.08$&$-1.13\pm 0.04$&$ 16.28\pm 1.08$\\
049&02&J080455.90+231501.8 & 2.162&1584-52943-0337&$ 1544.43\pm 107.60$&$-1.53\pm 0.03$&$ 15.41\pm 0.75$\\
050&01&J081213.95+431715.9 & 1.742&0546-52205-0403&$ 2010.89\pm 109.20$&$-1.68\pm 0.06$&$ 73.49\pm 0.70$\\
050&02&J081213.95+431715.9 & 1.742&0547-51959-0284&$ 2059.06\pm  94.30$&$-1.68\pm 0.06$&$ 75.13\pm 0.61$\\
051&01&J081822.63+434633.8 & 2.042&0547-51959-0122&$ 2322.17\pm 177.45$&$-1.28\pm 0.03$&$  2.78\pm 0.87$\\
051&02&J081822.63+434633.8 & 2.042&0547-52207-0157&$ 1937.76\pm 107.42$&$-1.26\pm 0.02$&$  8.13\pm 0.61$\\
052&01&J082238.64+420925.7 & 1.968&0761-52266-0244&$  670.99\pm  81.22$&$-0.69\pm 0.06$&$  7.56\pm 1.44$\\
052&02&J082238.64+420925.7 & 1.968&0761-54524-0279&$ 1100.38\pm 147.55$&$-1.81\pm 0.09$&$  3.73\pm 0.99$\\
053&01&J083925.61+045420.2 & 2.447&1187-52708-0137&$  949.11\pm 126.67$&$ 1.14\pm 0.10$&$ 42.31\pm 1.67$\\
053&02&J083925.61+045420.2 & 2.447&1188-52650-0390&$ 1058.86\pm 121.28$&$ 1.48\pm 0.10$&$ 41.67\pm 1.38$\\
054&01&J084255.92+223431.9 & 2.714&2084-53360-0502&$ 2138.52\pm 229.21$&$-1.02\pm 0.03$&$ 19.41\pm 1.17$\\
054&02&J084255.92+223431.9 & 2.714&3373-54940-0062&$ 2407.33\pm 134.97$&$-1.25\pm 0.02$&$ 12.08\pm 0.59$\\
055&01&J090241.07+571829.0 & 2.070&0483-51942-0521&$  708.01\pm 123.95$&$ 0.00\pm 0.06$&$ 15.65\pm 1.76$\\
055&02&J090241.07+571829.0 & 2.070&0483-51902-0554&$  692.91\pm  82.68$&$-0.05\pm 0.05$&$ 18.90\pm 1.26$\\
056&01&J091512.53+305014.9 & 1.983&1938-53379-0383&$ 1209.80\pm 108.77$&$-1.12\pm 0.04$&$  4.57\pm 0.96$\\
056&02&J091512.53+305014.9 & 1.983&2401-53768-0472&$ 1287.03\pm  62.89$&$-1.13\pm 0.02$&$  4.05\pm 0.52$\\
057&01&J092015.68+350040.5 & 1.916&1273-52993-0630&$ 1187.70\pm 107.89$&$-1.28\pm 0.06$&$ 27.79\pm 0.95$\\
057&02&J092015.68+350040.5 & 1.916&1274-52995-0284&$ 1090.35\pm  99.96$&$-1.34\pm 0.06$&$ 28.18\pm 0.96$\\
058&01&J092527.71+151416.9 & 1.968&2440-53818-0622&$ 1039.65\pm  96.15$&$-1.39\pm 0.05$&$ 14.58\pm 1.06$\\
058&02&J092527.71+151416.9 & 1.968&3192-54829-0504&$ 1268.01\pm  61.49$&$-1.47\pm 0.02$&$  8.45\pm 0.57$\\
059&01&J092720.29+101627.0 & 1.929&1740-53050-0065&$  983.36\pm 103.77$&$-2.25\pm 0.08$&$ 35.81\pm 1.37$\\
059&02&J092720.29+101627.0 & 1.929&3319-54915-0425&$  796.30\pm  43.66$&$-1.10\pm 0.04$&$ 34.60\pm 0.74$\\
060&01&J093300.21+544905.2 & 1.825&0556-51991-0313&$  925.17\pm 103.13$&$-1.09\pm 0.09$&$ 10.55\pm 1.34$\\
060&02&J093300.21+544905.2 & 1.825&0555-52266-0154&$  919.61\pm  84.46$&$-1.62\pm 0.08$&$  7.42\pm 1.05$\\
061&01&J093333.29+410522.8 & 2.153&0940-52670-0287&$ 2118.21\pm 226.18$&$-1.12\pm 0.04$&$  7.87\pm 1.08$\\
061&02&J093333.29+410522.8 & 2.153&0939-52636-0034&$ 2275.52\pm 138.07$&$-1.06\pm 0.02$&$  6.47\pm 0.65$\\
\hline
\end{tabular}
\end{table*}

\setcounter{table}{1}
\begin{table*}
\centering \caption{-- continue}
\begin{tabular}{lccccccccccccccccccc}

\hline \hline
  No.  & epoch &name  & z& Plate-MJD-Fiber &$L_{1500}$ & $\alpha$ & BAL EW \\
  (1) & (2) & (3) & (4) & (5) & (6) & (7) & (8)\\
\hline
062&01&J093548.50+363121.9 & 2.977&1275-52996-0145&$ 2526.87\pm 224.46$&$-1.42\pm 0.03$&$ 20.97\pm 1.04$\\
062&02&J093548.50+363121.9 & 2.977&3223-54865-0303&$ 2310.33\pm 159.98$&$-1.04\pm 0.02$&$ 33.81\pm 0.76$\\
063&01&J093620.52+004649.2 & 1.719&0476-52027-0442&$ 1099.22\pm 173.80$&$-1.34\pm 0.10$&$  4.75\pm 1.67$\\
063&02&J093620.52+004649.2 & 1.719&0476-52314-0444&$  772.49\pm 110.71$&$-0.91\pm 0.09$&$  6.41\pm 1.51$\\
064&01&J093859.27+150118.6 & 2.193&2581-54085-0228&$ 1379.40\pm 142.56$&$-0.83\pm 0.04$&$  4.16\pm 1.12$\\
064&02&J093859.27+150118.6 & 2.193&3196-54834-0640&$ 1802.93\pm  83.73$&$-0.74\pm 0.02$&$  1.28\pm 0.48$\\
065&01&J094338.21-010019.3 & 2.377&0266-51630-0124&$ 1639.51\pm 149.56$&$-0.77\pm 0.04$&$ 30.02\pm 0.97$\\
065&02&J094338.21-010019.3 & 2.377&0266-51602-0131&$ 1723.56\pm 150.00$&$-0.92\pm 0.03$&$ 31.13\pm 0.93$\\
066&01&J094425.46+610934.4 & 2.271&0486-51910-0120&$ 1121.96\pm 112.69$&$-0.55\pm 0.04$&$ 35.61\pm 1.08$\\
066&02&J094425.46+610934.4 & 2.271&2403-53795-0124&$ 1171.95\pm  60.84$&$-1.23\pm 0.02$&$ 32.36\pm 0.52$\\
067&01&J094456.75+544117.9 & 1.895&0769-54530-0327&$  785.66\pm  85.51$&$-0.60\pm 0.07$&$  6.59\pm 1.14$\\
067&02&J094456.75+544117.9 & 1.895&3169-54821-0377&$  698.54\pm  78.91$&$-0.29\pm 0.06$&$  5.57\pm 1.15$\\
068&01&J094602.23+380059.3 & 2.068&1276-53035-0160&$ 1675.13\pm 130.50$&$-1.23\pm 0.03$&$ 21.74\pm 0.83$\\
068&02&J094602.23+380059.3 & 2.068&3223-54865-0566&$ 1599.50\pm  76.83$&$-0.85\pm 0.02$&$ 23.75\pm 0.52$\\
069&01&J095357.00+040039.2 & 2.427&0571-52286-0560&$ 1310.91\pm 134.76$&$-0.93\pm 0.03$&$ 19.92\pm 1.12$\\
069&02&J095357.00+040039.2 & 2.427&0572-52289-0312&$ 1306.36\pm 127.44$&$-1.09\pm 0.03$&$ 20.38\pm 1.06$\\
070&01&J095901.24+550408.2 & 2.180&0945-52652-0436&$ 1194.36\pm  79.91$&$-0.89\pm 0.03$&$ 19.17\pm 0.87$\\
070&02&J095901.24+550408.2 & 2.180&3169-54821-0627&$ 1675.41\pm  71.33$&$-1.39\pm 0.02$&$ 12.73\pm 0.55$\\
071&01&J100109.51+133433.6 & 1.841&2584-54153-0020&$ 1428.41\pm 119.67$&$-0.71\pm 0.06$&$  4.96\pm 0.91$\\
071&02&J100109.51+133433.6 & 1.841&3248-54880-0120&$ 1554.03\pm  87.20$&$-0.97\pm 0.04$&$  5.13\pm 0.58$\\
072&01&J100318.99+521506.3 & 3.325&0903-52385-0177&$ 3897.33\pm 280.68$&$-1.70\pm 0.04$&$ 23.17\pm 0.87$\\
072&02&J100318.99+521506.3 & 3.325&0903-52400-0179&$ 3813.42\pm 235.67$&$-1.62\pm 0.03$&$ 22.77\pm 0.94$\\
073&01&J100619.31+625334.9 & 2.001&0487-51943-0077&$  899.42\pm  91.37$&$-1.05\pm 0.04$&$ 36.20\pm 1.08$\\
073&02&J100619.31+625334.9 & 2.001&3294-54918-0102&$  715.36\pm  67.37$&$-1.53\pm 0.05$&$ 30.09\pm 1.06$\\
074&01&J100716.69+030438.6 & 2.124&0501-52235-0606&$ 2689.48\pm 125.81$&$-1.23\pm 0.02$&$  6.36\pm 0.48$\\
074&02&J100716.69+030438.6 & 2.124&3257-54888-0474&$ 1927.77\pm  83.14$&$-0.91\pm 0.02$&$ 10.64\pm 0.44$\\
075&01&J100912.49+252055.2 & 2.205&2406-54084-0170&$  793.51\pm  60.54$&$-0.85\pm 0.03$&$ 34.54\pm 0.85$\\
075&02&J100912.49+252055.2 & 2.205&2347-53757-0213&$  861.28\pm 119.16$&$-0.71\pm 0.05$&$ 31.59\pm 1.54$\\
076&01&J101425.11+032003.7 & 2.145&0574-52347-0212&$ 3739.76\pm 177.15$&$-1.42\pm 0.02$&$  7.23\pm 0.51$\\
076&02&J101425.11+032003.7 & 2.145&0574-52366-0218&$ 4422.25\pm 214.56$&$-1.42\pm 0.02$&$  7.33\pm 0.48$\\
077&01&J101542.04+430455.6 & 2.420&1218-52709-0514&$ 5537.78\pm 244.90$&$-1.55\pm 0.01$&$  8.24\pm 0.46$\\
077&02&J101542.04+430455.6 & 2.420&3287-54941-0433&$ 4284.78\pm 154.82$&$-1.02\pm 0.01$&$ 12.99\pm 0.36$\\
078&01&J101616.34+383817.3 & 1.953&1427-52996-0479&$  853.61\pm  65.06$&$-0.89\pm 0.04$&$ 49.94\pm 0.88$\\
078&02&J101616.34+383817.3 & 1.953&3262-54884-0356&$ 1008.76\pm  62.58$&$-0.59\pm 0.04$&$ 54.69\pm 0.70$\\
079&01&J102106.78+303137.5 & 3.061&2351-53772-0409&$ 2114.50\pm 168.99$&$-1.97\pm 0.03$&$  7.78\pm 1.01$\\
079&02&J102106.78+303137.5 & 3.061&2351-53786-0409&$ 2376.23\pm 254.61$&$-2.01\pm 0.04$&$ 10.03\pm 1.23$\\
080&01&J102156.84+282735.8 & 1.880&2351-53772-0128&$  610.67\pm  77.81$&$-0.80\pm 0.09$&$ 26.23\pm 1.33$\\
080&02&J102156.84+282735.8 & 1.880&3260-54883-0624&$  542.65\pm  55.27$&$-0.93\pm 0.08$&$ 28.45\pm 1.04$\\
081&01&J102250.16+483631.1 & 2.069&0873-52347-0522&$ 1121.23\pm  91.69$&$-1.01\pm 0.04$&$ 31.33\pm 1.03$\\
081&02&J102250.16+483631.1 & 2.069&0873-52674-0555&$ 1373.20\pm  95.84$&$-1.43\pm 0.03$&$ 30.29\pm 0.87$\\
082&01&J102754.03+182221.6 & 3.078&2591-54140-0453&$ 2578.13\pm 231.53$&$-1.20\pm 0.03$&$ 27.76\pm 1.00$\\
082&02&J102754.03+182221.6 & 3.078&2868-54451-0518&$ 2370.06\pm  84.24$&$-1.07\pm 0.02$&$ 31.29\pm 0.46$\\
083&01&J103006.62+271325.9 & 1.734&2353-53794-0164&$  871.74\pm  99.58$&$-0.60\pm 0.08$&$ 41.39\pm 1.29$\\
083&02&J103006.62+271325.9 & 1.734&3261-54881-0582&$  893.54\pm  43.96$&$-0.94\pm 0.05$&$ 43.90\pm 0.60$\\
084&01&J104010.46+432811.6 & 2.595&2567-54179-0020&$ 2204.80\pm 105.51$&$-1.21\pm 0.02$&$ 20.61\pm 0.59$\\
084&02&J104010.46+432811.6 & 2.595&3258-54884-0219&$ 1708.79\pm 115.30$&$-0.98\pm 0.02$&$ 20.60\pm 0.83$\\
085&01&J104945.36+285823.3 & 2.154&2359-53800-0332&$  732.44\pm 101.18$&$-0.15\pm 0.05$&$ 42.83\pm 1.42$\\
085&02&J104945.36+285823.3 & 2.154&2359-53826-0332&$  764.55\pm 112.34$&$-0.18\pm 0.05$&$ 42.27\pm 1.48$\\
086&01&J105012.58+001158.8 & 2.206&2409-54210-0538&$  366.41\pm  38.80$&$-0.38\pm 0.04$&$  5.27\pm 1.19$\\
086&02&J105012.58+001158.8 & 2.206&2569-54234-0554&$  445.01\pm  48.59$&$-0.71\pm 0.04$&$  6.90\pm 1.21$\\
087&01&J105416.51+512326.0 & 2.340&0876-52669-0533&$ 1799.34\pm 142.16$&$-1.56\pm 0.03$&$ 10.81\pm 0.86$\\
087&02&J105416.51+512326.0 & 2.340&0876-52346-0535&$ 1548.09\pm 107.72$&$-1.30\pm 0.02$&$  9.26\pm 0.78$\\
088&01&J105657.54+492957.9 & 2.162&0876-52669-0014&$ 2073.65\pm 172.16$&$-1.03\pm 0.03$&$  1.91\pm 0.88$\\
088&02&J105657.54+492957.9 & 2.162&0876-52346-0017&$ 1211.93\pm 138.04$&$-1.67\pm 0.04$&$  0.97\pm 1.20$\\
089&01&J110015.55+271451.7 & 3.339&2359-53800-0003&$ 3164.53\pm 191.43$&$-1.22\pm 0.04$&$ 22.14\pm 0.79$\\
089&02&J110015.55+271451.7 & 3.339&2359-53826-0003&$ 3247.59\pm 170.00$&$-1.19\pm 0.04$&$ 21.17\pm 0.82$\\
090&01&J110152.91+275838.6 & 2.865&2211-53786-0256&$ 2034.26\pm 323.16$&$-0.26\pm 0.04$&$  8.94\pm 1.42$\\
090&02&J110152.91+275838.6 & 2.865&2870-54534-0165&$ 2008.90\pm 132.62$&$-0.11\pm 0.02$&$  6.97\pm 0.72$\\
091&01&J110208.59+660156.5 & 2.064&0490-51929-0142&$  897.03\pm  75.33$&$-1.14\pm 0.04$&$ 24.04\pm 0.89$\\
091&02&J110208.59+660156.5 & 2.064&3171-54862-0164&$ 1034.75\pm  68.61$&$-1.04\pm 0.03$&$ 18.57\pm 0.69$\\
092&01&J110427.08+054848.3 & 3.006&0581-52353-0328&$ 2714.96\pm 233.04$&$-2.29\pm 0.03$&$  4.70\pm 1.12$\\
092&02&J110427.08+054848.3 & 3.006&0581-52356-0328&$ 2849.43\pm 230.02$&$-2.23\pm 0.03$&$  4.87\pm 1.03$\\

\hline
\end{tabular}
\end{table*}

\setcounter{table}{1}
\begin{table*}
\centering \caption{-- continue}
\begin{tabular}{lccccccccccccccccccc}

\hline \hline
  No.  & epoch &name  & z& Plate-MJD-Fiber &$L_{1500}$ & $\alpha$ & BAL EW \\
  (1) & (2) & (3) & (4) & (5) & (6) & (7) & (8)\\
\hline
093&01&J111313.29+102212.4 & 2.247&1222-52763-0322&$ 4914.87\pm 252.26$&$-0.99\pm 0.02$&$  8.25\pm 0.53$\\
093&02&J111313.29+102212.4 & 2.247&2393-54156-0420&$ 6065.31\pm 123.81$&$-1.08\pm 0.01$&$  7.90\pm 0.22$\\
094&01&J111651.98+463508.6 & 1.888&3216-54908-0385&$  540.64\pm  62.58$&$-1.52\pm 0.08$&$ 14.73\pm 1.18$\\
094&02&J111651.98+463508.6 & 1.888&3216-54853-0399&$  581.58\pm  51.68$&$-1.41\pm 0.06$&$ 11.95\pm 0.98$\\
095&01&J112239.20+602012.3 & 2.227&0951-52398-0523&$ 1142.37\pm 128.40$&$-1.50\pm 0.04$&$  9.10\pm 1.17$\\
095&02&J112239.20+602012.3 & 2.227&3328-54964-0081&$  966.08\pm  94.23$&$-1.73\pm 0.03$&$  5.71\pm 0.93$\\
096&01&J112258.77+164540.3 & 3.031&2499-54176-0308&$ 8025.55\pm 284.56$&$-1.04\pm 0.03$&$ 42.12\pm 0.45$\\
096&02&J112258.77+164540.3 & 3.031&3327-54951-0153&$ 7874.83\pm 162.16$&$-0.81\pm 0.03$&$ 39.93\pm 0.30$\\
097&01&J112703.06+450516.4 & 1.857&1366-53063-0321&$ 1130.13\pm 113.64$&$-1.85\pm 0.10$&$ 20.09\pm 1.06$\\
097&02&J112703.06+450516.4 & 1.857&3215-54861-0399&$  671.43\pm  50.17$&$-1.64\pm 0.07$&$ 12.46\pm 0.78$\\
098&01&J112939.84+600728.9 & 1.724&0952-52409-0412&$ 1091.60\pm 108.85$&$-1.44\pm 0.07$&$  5.73\pm 1.11$\\
098&02&J112939.84+600728.9 & 1.724&3211-54852-0424&$ 1024.48\pm  71.53$&$-1.71\pm 0.05$&$  2.00\pm 0.77$\\
099&01&J113529.56+584803.8 & 1.781&0952-52409-0150&$  778.03\pm  98.26$&$-1.23\pm 0.09$&$ 21.69\pm 1.36$\\
099&02&J113529.56+584803.8 & 1.781&2881-54502-0296&$  672.88\pm  82.21$&$-1.60\pm 0.09$&$ 30.81\pm 1.37$\\
100&01&J113621.05+005021.2 & 3.428&0282-51630-0535&$ 8531.06\pm 370.79$&$-1.77\pm 0.02$&$  6.83\pm 0.46$\\
100&02&J113621.05+005021.2 & 3.428&0282-51658-0535&$ 8891.49\pm 282.79$&$-1.71\pm 0.02$&$  6.31\pm 0.38$\\
101&01&J115944.82+011206.9 & 2.002&0285-51663-0530&$ 5261.82\pm 266.65$&$-2.11\pm 0.02$&$ 18.10\pm 0.53$\\
101&02&J115944.82+011206.9 & 2.002&0285-51930-0540&$ 4932.98\pm 191.60$&$-1.67\pm 0.02$&$ 18.12\pm 0.40$\\
102&01&J120217.29+332108.7 & 2.252&2089-53498-0292&$ 1772.50\pm 118.41$&$-1.64\pm 0.02$&$  8.47\pm 0.71$\\
102&02&J120217.29+332108.7 & 2.252&2095-53474-0448&$ 1800.74\pm 181.35$&$-1.57\pm 0.03$&$  9.41\pm 0.97$\\
103&01&J120653.39+492919.3 & 1.845&0969-52442-0107&$ 2508.88\pm 130.53$&$-1.71\pm 0.05$&$ 64.15\pm 0.55$\\
103&02&J120653.39+492919.3 & 1.845&2919-54537-0170&$ 2198.63\pm  58.20$&$-1.55\pm 0.03$&$ 59.46\pm 0.28$\\
104&01&J120822.25+302423.9 & 2.252&2230-53799-0353&$ 2827.30\pm 216.24$&$-1.43\pm 0.02$&$  5.31\pm 0.73$\\
104&02&J120822.25+302423.9 & 2.252&3181-54860-0025&$ 2552.20\pm 118.31$&$-1.21\pm 0.01$&$  1.68\pm 0.49$\\
105&01&J121147.38+203402.4 & 2.413&2610-54476-0341&$ 2679.76\pm 197.05$&$-1.69\pm 0.03$&$ 23.20\pm 0.82$\\
105&02&J121147.38+203402.4 & 2.413&2918-54554-0622&$ 2761.27\pm  93.74$&$-1.76\pm 0.01$&$ 17.27\pm 0.39$\\
106&01&J121328.78-025617.8 & 2.153&0332-52367-0065&$ 2299.24\pm 184.48$&$-1.11\pm 0.03$&$ 13.76\pm 0.85$\\
106&02&J121328.78-025617.8 & 2.153&0333-52313-0315&$ 2302.60\pm 149.92$&$-1.19\pm 0.03$&$ 11.93\pm 0.73$\\
107&01&J122604.28+034317.8 & 1.767&0519-52283-0565&$ 1956.38\pm 109.69$&$-1.37\pm 0.05$&$ 20.78\pm 0.58$\\
107&02&J122604.28+034317.8 & 1.767&3253-54941-0202&$ 2113.96\pm 101.23$&$-1.31\pm 0.03$&$ 21.63\pm 0.50$\\
108&01&J122951.77+351929.9 & 1.821&2010-53495-0309&$ 1731.18\pm 116.99$&$-1.42\pm 0.06$&$ 18.34\pm 0.79$\\
108&02&J122951.77+351929.9 & 1.821&3395-55004-0278&$ 1623.91\pm  80.70$&$-1.66\pm 0.04$&$ 14.79\pm 0.55$\\
109&01&J123303.50+620915.9 & 1.839&0780-52370-0109&$ 1419.77\pm 122.12$&$-1.45\pm 0.07$&$ 18.91\pm 0.89$\\
109&02&J123303.50+620915.9 & 1.839&0781-52373-0278&$ 1502.92\pm 152.26$&$-1.52\pm 0.08$&$ 19.42\pm 1.08$\\
110&01&J123411.74+615832.5 & 1.949&0780-52370-0065&$ 1047.71\pm 127.36$&$-1.04\pm 0.07$&$ 40.59\pm 1.32$\\
110&02&J123411.74+615832.5 & 1.949&0781-52373-0275&$  958.88\pm 113.42$&$-0.73\pm 0.07$&$ 38.02\pm 1.29$\\
111&01&J123736.42+143640.1 & 2.704&1768-53442-0038&$ 2075.61\pm 219.11$&$-1.00\pm 0.03$&$ 31.35\pm 1.25$\\
111&02&J123736.42+143640.1 & 2.704&3254-54889-0548&$ 2426.53\pm 129.76$&$-1.23\pm 0.02$&$ 31.44\pm 0.63$\\
112&01&J124140.08+131746.1 & 2.143&1694-53472-0336&$ 1303.73\pm 120.77$&$-1.37\pm 0.04$&$ 10.71\pm 1.07$\\
112&02&J124140.08+131746.1 & 2.143&3255-54885-0205&$ 1096.03\pm  72.40$&$-1.21\pm 0.02$&$ 13.47\pm 0.65$\\
113&01&J124551.44+010505.0 & 2.809&0291-51660-0607&$ 4162.68\pm 267.66$&$-1.54\pm 0.02$&$ 29.94\pm 0.92$\\
113&02&J124551.44+010505.0 & 2.809&0291-51928-0612&$ 3921.53\pm 236.91$&$-1.61\pm 0.02$&$ 29.11\pm 0.78$\\
114&01&J125950.76+183236.1 & 2.248&2616-54499-0124&$ 1083.87\pm 122.29$&$-1.04\pm 0.04$&$  3.88\pm 1.22$\\
114&02&J125950.76+183236.1 & 2.248&2924-54582-0083&$ 1031.82\pm  71.36$&$-0.92\pm 0.02$&$  2.58\pm 0.76$\\
115&01&J130136.12+000157.9 & 1.783&0293-51994-0074&$ 3457.46\pm 161.29$&$-1.54\pm 0.05$&$ 35.63\pm 0.56$\\
115&02&J130136.12+000157.9 & 1.783&0293-51689-0079&$ 3261.62\pm 129.30$&$-1.72\pm 0.04$&$ 35.73\pm 0.47$\\
116&01&J130221.80-004638.2 & 2.704&0293-51689-0012&$ 2020.28\pm 259.70$&$-0.83\pm 0.04$&$ 19.90\pm 1.76$\\
116&02&J130221.80-004638.2 & 2.704&0293-51994-0015&$ 1901.44\pm 217.58$&$-0.92\pm 0.04$&$ 14.52\pm 1.74$\\
117&01&J131416.43-015020.9 & 2.180&0340-51691-0508&$ 1280.93\pm 155.59$&$-1.65\pm 0.04$&$  7.17\pm 1.30$\\
117&02&J131416.43-015020.9 & 2.180&0340-51990-0514&$ 1386.06\pm 116.92$&$-1.86\pm 0.03$&$  5.04\pm 0.93$\\
118&01&J131433.83+032321.9 & 2.255&0525-52029-0572&$ 2105.90\pm 201.20$&$-1.75\pm 0.03$&$  5.48\pm 1.01$\\
118&02&J131433.83+032321.9 & 2.255&0525-52295-0576&$ 1720.59\pm 137.72$&$-1.46\pm 0.02$&$ 11.15\pm 0.81$\\
119&01&J131505.89+590157.5 & 1.932&0958-52410-0157&$ 2256.73\pm 137.19$&$-0.90\pm 0.04$&$ 25.40\pm 0.71$\\
119&02&J131505.89+590157.5 & 1.932&3237-54883-0067&$ 1889.68\pm  78.10$&$-0.63\pm 0.02$&$ 20.72\pm 0.46$\\
120&01&J131714.21+010013.0 & 2.698&0296-51578-0327&$ 3857.85\pm 267.72$&$-1.61\pm 0.03$&$ 23.92\pm 0.88$\\
120&02&J131714.21+010013.0 & 2.698&0296-51984-0329&$ 3796.38\pm 215.64$&$-1.55\pm 0.02$&$ 26.54\pm 0.76$\\
121&01&J131853.45+002211.5 & 2.074&0296-51578-0383&$ 1530.04\pm 143.96$&$-1.60\pm 0.04$&$ 19.34\pm 1.00$\\
121&02&J131853.45+002211.5 & 2.074&0296-51984-0390&$ 1372.49\pm 130.93$&$-1.47\pm 0.03$&$ 21.71\pm 0.99$\\
122&01&J131905.95+660415.7 & 1.710&0496-51973-0260&$ 1544.80\pm 169.04$&$-2.24\pm 0.08$&$  9.63\pm 1.14$\\
122&02&J131905.95+660415.7 & 1.710&0496-51988-0260&$ 1415.09\pm 161.92$&$-1.78\pm 0.07$&$  7.41\pm 1.15$\\
123&01&J132304.58-003856.5 & 1.827&0296-51578-0074&$ 1363.05\pm 108.92$&$-1.36\pm 0.07$&$ 25.33\pm 1.04$\\
123&02&J132304.58-003856.5 & 1.827&0296-51984-0076&$ 1328.15\pm 113.27$&$-1.11\pm 0.07$&$ 22.89\pm 1.08$\\
\hline
\end{tabular}
\end{table*}

\setcounter{table}{1}
\begin{table*}
\centering \caption{-- continue}
\begin{tabular}{lccccccccccccccccccc}

\hline \hline
  No.  & epoch &name  & z& Plate-MJD-Fiber &$L_{1500}$ & $\alpha$ & BAL EW \\
  (1) & (2) & (3) & (4) & (5) & (6) & (7) & (8)\\
\hline

124&01&J132422.54+245222.4 & 2.363&2664-54524-0537&$ 3194.67\pm 168.12$&$-1.38\pm 0.02$&$ 28.43\pm 0.56$\\
124&02&J132422.54+245222.4 & 2.363&3303-54950-0599&$ 2773.42\pm 102.74$&$-1.36\pm 0.01$&$ 27.57\pm 0.37$\\
125&01&J132827.06+581836.8 & 3.139&0960-52466-0304&$ 2794.53\pm 251.42$&$-1.76\pm 0.05$&$ 11.45\pm 1.38$\\
125&02&J132827.06+581836.8 & 3.139&0960-52425-0318&$ 2827.13\pm 188.18$&$-2.15\pm 0.04$&$ 14.99\pm 0.87$\\
126&01&J133138.50+004221.1 & 2.424&0298-51955-0374&$ 2264.05\pm 164.18$&$-1.82\pm 0.02$&$ 11.14\pm 0.78$\\
126&02&J133138.50+004221.1 & 2.424&0298-51662-0376&$ 2168.88\pm 208.04$&$-1.62\pm 0.03$&$  6.24\pm 1.10$\\
127&01&J133514.39+531805.8 & 1.874&1041-52724-0039&$ 1578.48\pm 127.97$&$-1.17\pm 0.06$&$ 13.08\pm 0.91$\\
127&02&J133514.39+531805.8 & 1.874&3318-54951-0115&$ 1395.43\pm  83.67$&$-0.99\pm 0.04$&$ 14.02\pm 0.63$\\
128&01&J134101.28+083755.7 & 2.494&1804-53886-0318&$ 1512.42\pm 144.88$&$ 0.30\pm 0.08$&$ 53.75\pm 1.08$\\
128&02&J134101.28+083755.7 & 2.494&2928-54614-0128&$ 1427.07\pm  68.11$&$-0.06\pm 0.05$&$ 56.01\pm 0.52$\\
129&01&J134544.55+002810.7 & 2.453&0300-51943-0382&$ 1918.11\pm 151.67$&$-1.34\pm 0.03$&$ 20.84\pm 0.90$\\
129&02&J134544.55+002810.7 & 2.453&0300-51666-0426&$ 2153.34\pm 160.97$&$-1.65\pm 0.03$&$ 17.78\pm 0.83$\\
130&01&J135448.04+501137.9 & 2.141&1670-54553-0483&$  722.58\pm  85.80$&$-0.57\pm 0.05$&$ 11.40\pm 1.31$\\
130&02&J135448.04+501137.9 & 2.141&1670-53438-0486&$  771.58\pm  81.16$&$-0.76\pm 0.04$&$  5.88\pm 1.16$\\
131&01&J135559.03-002413.6 & 2.337&0301-51641-0266&$ 3095.13\pm 179.97$&$-1.68\pm 0.02$&$ 14.60\pm 0.65$\\
131&02&J135559.03-002413.6 & 2.337&0301-51942-0267&$ 2769.95\pm 162.58$&$-1.72\pm 0.02$&$ 15.86\pm 0.67$\\
132&01&J135721.77+005501.1 & 1.997&0301-51942-0408&$ 2278.58\pm 128.98$&$-1.27\pm 0.02$&$  8.54\pm 0.59$\\
132&02&J135721.77+005501.1 & 1.997&0301-51641-0411&$ 2084.64\pm 133.82$&$-1.31\pm 0.03$&$  8.32\pm 0.66$\\
133&01&J135941.58+000851.9 & 1.736&0301-51942-0517&$  862.52\pm  97.91$&$-1.72\pm 0.08$&$  5.32\pm 1.17$\\
133&02&J135941.58+000851.9 & 1.736&0301-51641-0520&$  857.03\pm  94.83$&$-1.39\pm 0.09$&$  7.58\pm 1.28$\\
134&01&J142333.56+573909.5 & 1.870&0789-52342-0229&$  789.71\pm 102.50$&$-0.73\pm 0.10$&$ 25.77\pm 1.49$\\
134&02&J142333.56+573909.5 & 1.870&2547-53917-0265&$  735.62\pm 110.03$&$-0.08\pm 0.11$&$ 24.84\pm 1.74$\\
135&01&J142910.55+634603.6 & 1.886&0499-51988-0143&$  686.52\pm  76.90$&$-1.47\pm 0.08$&$  9.11\pm 1.24$\\
135&02&J142910.55+634603.6 & 1.886&2947-54533-0327&$  522.58\pm  63.00$&$-1.34\pm 0.09$&$ 10.04\pm 1.34$\\
136&01&J143117.07+632701.7 & 1.891&0499-51988-0105&$ 1828.02\pm  94.61$&$-1.04\pm 0.03$&$ 15.91\pm 0.54$\\
136&02&J143117.07+632701.7 & 1.891&2947-54533-0370&$ 1876.33\pm 135.72$&$-1.41\pm 0.05$&$ 11.06\pm 0.75$\\
137&01&J143130.03+570138.8 & 1.797&0790-52441-0245&$ 2142.59\pm 140.70$&$-1.74\pm 0.06$&$ 23.75\pm 0.80$\\
137&02&J143130.03+570138.8 & 1.797&0790-52346-0257&$ 2224.85\pm 146.92$&$-1.96\pm 0.06$&$ 25.38\pm 0.75$\\
138&01&J143307.40+003319.0 & 2.744&0306-51637-0546&$ 1573.84\pm 179.18$&$-1.87\pm 0.04$&$ 17.83\pm 1.57$\\
138&02&J143307.40+003319.0 & 2.744&0306-51690-0558&$ 1552.12\pm 176.33$&$-1.77\pm 0.04$&$ 17.49\pm 1.84$\\
139&01&J143612.69+443812.6 & 1.848&1288-52731-0105&$  896.37\pm 125.93$&$-0.62\pm 0.10$&$ 18.20\pm 1.50$\\
139&02&J143612.69+443812.6 & 1.848&1289-52734-0350&$  930.19\pm 111.76$&$-0.51\pm 0.09$&$ 14.97\pm 1.30$\\
140&01&J143641.24+001558.9 & 1.867&0306-51637-0628&$ 1072.35\pm  87.18$&$-1.19\pm 0.08$&$ 28.91\pm 0.97$\\
140&02&J143641.24+001558.9 & 1.867&0306-51690-0629&$ 1058.77\pm 106.35$&$-1.32\pm 0.09$&$ 31.47\pm 1.17$\\
141&01&J143758.06+011119.5 & 2.045&0307-51663-0443&$ 1022.18\pm 114.78$&$-0.66\pm 0.04$&$ 24.68\pm 1.18$\\
141&02&J143758.06+011119.5 & 2.045&0536-52024-0217&$ 1014.26\pm 113.98$&$-0.80\pm 0.04$&$ 26.18\pm 1.20$\\
142&01&J143907.51-010616.7 & 1.819&0307-51663-0089&$  539.73\pm 116.89$&$-0.27\pm 0.14$&$ 12.58\pm 2.53$\\
142&02&J143907.51-010616.7 & 1.819&0919-52409-0566&$  674.14\pm 115.50$&$-0.35\pm 0.11$&$ 10.88\pm 1.90$\\
143&01&J144136.54+632519.4 & 1.779&0609-52339-0322&$ 1067.87\pm 136.16$&$-1.70\pm 0.10$&$ 13.63\pm 1.38$\\
143&02&J144136.54+632519.4 & 1.779&2947-54533-0153&$  730.82\pm  76.55$&$-2.36\pm 0.08$&$  7.84\pm 1.23$\\
144&01&J144351.38+560325.6 & 2.275&0791-52347-0247&$ 1532.55\pm 145.80$&$-1.60\pm 0.03$&$ 11.38\pm 1.07$\\
144&02&J144351.38+560325.6 & 2.275&0791-52435-0247&$ 1632.08\pm 169.46$&$-1.62\pm 0.04$&$ 10.54\pm 1.15$\\
145&01&J144412.36+582636.9 & 2.336&0790-52441-0583&$ 1187.20\pm 149.14$&$-0.58\pm 0.04$&$  2.63\pm 1.32$\\
145&02&J144412.36+582636.9 & 2.336&0790-52346-0584&$ 1329.29\pm 158.70$&$-0.95\pm 0.04$&$  5.10\pm 1.31$\\
146&01&J144514.86-002358.1 & 2.237&2934-54626-0223&$ 3403.12\pm  95.96$&$ 0.21\pm 0.02$&$ 24.90\pm 0.27$\\
146&02&J144514.86-002358.1 & 2.237&2909-54653-0222&$ 3109.81\pm 142.56$&$ 0.21\pm 0.02$&$ 24.98\pm 0.48$\\
147&01&J144935.96+631836.0 & 1.735&0609-52339-0538&$  891.96\pm 130.81$&$-1.12\pm 0.10$&$ 17.81\pm 1.66$\\
147&02&J144935.96+631836.0 & 1.735&2947-54533-0044&$  615.93\pm  76.12$&$-1.45\pm 0.08$&$ 16.42\pm 1.51$\\
148&01&J145110.68+040925.2 & 1.732&0588-52029-0440&$ 1295.74\pm 167.75$&$-1.52\pm 0.09$&$ 19.43\pm 1.37$\\
148&02&J145110.68+040925.2 & 1.732&0588-52045-0440&$ 1408.12\pm 120.28$&$-1.66\pm 0.07$&$ 19.36\pm 0.93$\\
149&01&J145353.45+040124.0 & 2.088&0588-52029-0488&$ 1917.34\pm 147.12$&$-1.69\pm 0.03$&$ 10.69\pm 0.99$\\
149&02&J145353.45+040124.0 & 2.088&0588-52045-0496&$ 2106.68\pm 102.42$&$-1.82\pm 0.02$&$ 10.19\pm 0.70$\\
150&01&J145943.02+010601.5 & 2.092&0310-51616-0393&$ 2071.46\pm 188.94$&$-0.97\pm 0.03$&$  2.97\pm 0.97$\\
150&02&J145943.02+010601.5 & 2.092&0538-52029-0012&$ 2210.66\pm 160.70$&$-1.58\pm 0.03$&$  3.39\pm 0.79$\\
151&01&J150033.52+003353.6 & 2.438&0310-51616-0363&$ 3455.47\pm 259.64$&$-1.11\pm 0.02$&$ 12.07\pm 0.78$\\
151&02&J150033.52+003353.6 & 2.438&0310-51990-0388&$ 2449.34\pm 197.15$&$-0.53\pm 0.03$&$ 18.26\pm 0.83$\\
152&01&J150332.93+440120.6 & 2.049&1676-53147-0028&$ 1118.21\pm 104.63$&$-0.68\pm 0.04$&$ 16.02\pm 1.19$\\
152&02&J150332.93+440120.6 & 2.049&1677-53148-0358&$ 1205.95\pm  91.43$&$-0.83\pm 0.04$&$ 18.93\pm 1.02$\\
153&01&J150428.59-002015.9 & 1.865&0310-51616-0170&$ 1359.99\pm 154.39$&$-1.30\pm 0.09$&$  9.28\pm 1.28$\\
153&02&J150428.59-002015.9 & 1.865&0310-51990-0199&$ 1261.01\pm 131.14$&$-1.34\pm 0.08$&$  9.30\pm 1.12$\\
154&01&J150659.88+420652.7 & 1.868&1291-52735-0489&$ 1156.12\pm 101.45$&$-1.54\pm 0.08$&$ 14.99\pm 1.00$\\
154&02&J150659.88+420652.7 & 1.868&1291-52738-0489&$ 1187.00\pm 112.53$&$-1.65\pm 0.08$&$ 16.55\pm 1.03$\\

\hline
\end{tabular}
\end{table*}

\setcounter{table}{1}
\begin{table*}
\centering \caption{-- continue}
\begin{tabular}{lccccccccccccccccccc}

\hline \hline
  No.  & epoch &name  & z& Plate-MJD-Fiber &$L_{1500}$ & $\alpha$ & BAL EW \\
  (1) & (2) & (3) & (4) & (5) & (6) & (7) & (8)\\
\hline

155&01&J152057.81+461641.3 & 1.962&1050-52721-0629&$  648.26\pm 113.38$&$ 0.72\pm 0.07$&$ 55.97\pm 1.92$\\
155&02&J152057.81+461641.3 & 1.962&1331-52766-0248&$  630.92\pm  91.87$&$ 0.39\pm 0.08$&$ 60.70\pm 1.73$\\
156&01&J152149.78+010236.4 & 2.231&0313-51673-0339&$ 1365.55\pm 164.68$&$-1.48\pm 0.04$&$  6.09\pm 1.25$\\
156&02&J152149.78+010236.4 & 2.231&2953-54560-0209&$ 2686.50\pm 183.33$&$-1.83\pm 0.02$&$  3.84\pm 0.70$\\
157&01&J153045.76+383952.3 & 2.022&1293-52765-0068&$ 2835.80\pm 153.79$&$-1.74\pm 0.02$&$ 16.74\pm 0.54$\\
157&02&J153045.76+383952.3 & 2.022&1294-52753-0329&$ 2418.72\pm 167.68$&$-1.36\pm 0.03$&$ 14.27\pm 0.75$\\
158&01&J153715.74+582933.9 & 2.590&0615-52345-0577&$ 6004.06\pm 393.01$&$-0.68\pm 0.02$&$ 16.86\pm 0.59$\\
158&02&J153715.74+582933.9 & 2.590&0615-52347-0585&$ 6259.21\pm 269.68$&$-0.75\pm 0.01$&$ 16.87\pm 0.42$\\
159&01&J154359.43+535903.1 & 2.371&0616-52374-0097&$ 6736.25\pm 391.38$&$-1.73\pm 0.02$&$  4.99\pm 0.68$\\
159&02&J154359.43+535903.1 & 2.371&0616-52442-0120&$ 6930.08\pm 274.79$&$-1.56\pm 0.01$&$  4.20\pm 0.41$\\
160&01&J160649.23+451051.6 & 2.826&0814-52370-0355&$ 2815.78\pm 261.00$&$-0.58\pm 0.03$&$ 17.28\pm 1.31$\\
160&02&J160649.23+451051.6 & 2.826&0814-52443-0359&$ 2792.49\pm 275.76$&$-0.60\pm 0.03$&$ 18.92\pm 1.31$\\
161&01&J163023.69+390841.7 & 2.008&1172-52759-0008&$  881.66\pm 139.10$&$-0.03\pm 0.06$&$ 14.28\pm 1.58$\\
161&02&J163023.69+390841.7 & 2.008&1173-52790-0453&$  790.73\pm 134.68$&$ 0.09\pm 0.07$&$ 12.60\pm 1.91$\\
162&01&J164419.97+274447.0 & 3.068&1690-53475-0350&$ 5126.54\pm 224.13$&$-1.39\pm 0.02$&$ 24.99\pm 0.59$\\
162&02&J164419.97+274447.0 & 3.068&2949-54557-0603&$ 4163.99\pm 231.63$&$-1.70\pm 0.02$&$ 27.11\pm 0.65$\\
163&01&J164424.98+274136.5 & 3.893&1690-53475-0342&$ 9125.81\pm 840.91$&$-1.07\pm 0.03$&$  5.79\pm 0.57$\\
163&02&J164424.98+274136.5 & 3.893&2949-54557-0602&$ 7408.34\pm 746.56$&$-1.33\pm 0.03$&$  4.58\pm 0.60$\\
164&01&J165443.30+372008.0 & 1.928&0632-52071-0125&$ 1332.06\pm 136.03$&$-1.58\pm 0.05$&$  3.47\pm 1.06$\\
164&02&J165443.30+372008.0 & 1.928&0820-52438-0371&$ 1157.64\pm  93.21$&$-1.79\pm 0.05$&$  3.72\pm 0.89$\\
165&01&J165631.20+353259.0 & 2.039&0819-52409-0006&$  857.54\pm 126.85$&$-0.30\pm 0.06$&$ 32.15\pm 1.58$\\
165&02&J165631.20+353259.0 & 2.039&0820-52438-0092&$ 1036.77\pm  87.36$&$-0.80\pm 0.04$&$ 35.35\pm 0.94$\\
166&01&J170105.08+372347.3 & 1.959&0820-52438-0614&$  796.09\pm 108.81$&$-0.73\pm 0.06$&$ 18.23\pm 1.29$\\
166&02&J170105.08+372347.3 & 1.959&0820-52405-0638&$  838.88\pm  97.70$&$-1.10\pm 0.06$&$ 16.51\pm 1.26$\\
167&01&J170633.06+615715.1 & 2.012&0351-51780-0555&$ 1272.88\pm 127.51$&$-1.76\pm 0.04$&$  9.93\pm 1.12$\\
167&02&J170633.06+615715.1 & 2.012&0351-51695-0556&$ 1321.14\pm 137.66$&$-1.64\pm 0.04$&$ 11.44\pm 0.98$\\
168&01&J170931.00+630357.2 & 2.395&0352-51789-0310&$ 3799.17\pm 193.13$&$-1.20\pm 0.02$&$ 26.52\pm 0.58$\\
168&02&J170931.00+630357.2 & 2.395&0352-51694-0311&$ 3023.91\pm 174.46$&$-0.91\pm 0.03$&$ 29.68\pm 0.67$\\
169&01&J171731.04+621912.0 & 2.118&0352-51789-0209&$ 1199.22\pm 122.94$&$-1.53\pm 0.04$&$ 10.12\pm 1.11$\\
169&02&J171731.04+621912.0 & 2.118&0352-51694-0215&$ 1113.01\pm 138.59$&$-1.66\pm 0.05$&$  9.99\pm 1.33$\\
170&01&J172012.40+545601.1 & 2.100&0357-51813-0189&$ 1923.35\pm 103.37$&$-1.78\pm 0.02$&$ 14.19\pm 0.59$\\
170&02&J172012.40+545601.1 & 2.100&0367-51997-0184&$ 2071.91\pm 100.37$&$-2.02\pm 0.02$&$ 14.55\pm 0.49$\\
171&01&J173802.90+535047.2 & 1.918&0362-51999-0134&$  989.13\pm 113.87$&$-0.35\pm 0.06$&$  6.12\pm 1.21$\\
171&02&J173802.90+535047.2 & 1.918&0360-51780-0134&$  703.79\pm 148.22$&$ 0.70\pm 0.10$&$  2.17\pm 2.26$\\
172&01&J212127.78-081737.5 & 1.906&0640-52178-0286&$ 1276.19\pm 169.68$&$-0.84\pm 0.06$&$  2.42\pm 1.31$\\
172&02&J212127.78-081737.5 & 1.906&0640-52200-0295&$ 1189.13\pm 180.42$&$-0.83\pm 0.08$&$  1.95\pm 1.45$\\
173&01&J213113.93-083913.5 & 1.983&0641-52176-0297&$  632.51\pm 115.90$&$-0.33\pm 0.08$&$ 22.50\pm 1.99$\\
173&02&J213113.93-083913.5 & 1.983&0641-52199-0299&$  732.44\pm 103.17$&$-0.65\pm 0.07$&$ 22.87\pm 1.63$\\
174&01&J213138.07-002537.8 & 1.837&0989-52468-0273&$  591.30\pm 115.95$&$-0.25\pm 0.13$&$ 34.51\pm 2.15$\\
174&02&J213138.07-002537.8 & 1.837&1963-54331-0238&$  301.29\pm  44.97$&$ 0.36\pm 0.08$&$ 17.94\pm 1.60$\\
175&01&J213138.93-070013.4 & 2.048&0641-52176-0386&$ 1272.71\pm 116.96$&$-1.55\pm 0.04$&$ 29.72\pm 1.07$\\
175&02&J213138.93-070013.4 & 2.048&0641-52199-0390&$ 1258.61\pm 101.25$&$-1.46\pm 0.04$&$ 32.74\pm 0.97$\\
176&01&J213508.59-075502.1 & 1.799&0641-52176-0108&$ 1611.66\pm 128.10$&$-1.45\pm 0.06$&$  6.42\pm 0.94$\\
176&02&J213508.59-075502.1 & 1.799&0641-52199-0116&$ 1778.34\pm 140.06$&$-1.71\pm 0.06$&$  8.97\pm 0.93$\\
177&01&J213648.17-001546.6 & 2.180&0989-52468-0104&$ 1169.71\pm 128.27$&$-1.46\pm 0.04$&$ 13.46\pm 1.23$\\
177&02&J213648.17-001546.6 & 2.180&1152-52941-0317&$ 1320.84\pm 103.66$&$-1.42\pm 0.03$&$  8.75\pm 0.83$\\
178&01&J221555.98+010127.0 & 2.241&1104-52912-0530&$ 1838.20\pm 148.25$&$-1.66\pm 0.03$&$ 13.56\pm 0.87$\\
178&02&J221555.98+010127.0 & 2.241&3146-54773-0609&$ 1671.33\pm 105.68$&$-1.21\pm 0.02$&$  8.65\pm 0.66$\\
179&01&J222107.29+125627.2 & 1.732&0736-52210-0027&$ 1291.86\pm 161.31$&$-1.06\pm 0.08$&$ 24.85\pm 1.38$\\
179&02&J222107.29+125627.2 & 1.732&0736-52221-0031&$ 1514.98\pm 151.15$&$-1.48\pm 0.07$&$ 23.80\pm 1.06$\\
180&01&J224248.92-085822.8 & 2.390&0722-52224-0075&$ 1594.81\pm 161.98$&$-1.80\pm 0.03$&$ 11.72\pm 1.03$\\
180&02&J224248.92-085822.8 & 2.390&0723-52201-0358&$ 1654.45\pm 160.01$&$-1.78\pm 0.03$&$ 11.31\pm 0.96$\\
181&01&J224324.59-005330.7 & 1.895&0378-52146-0216&$ 1001.88\pm 150.58$&$-0.65\pm 0.09$&$ 19.33\pm 1.58$\\
181&02&J224324.59-005330.7 & 1.895&0675-52590-0208&$ 1159.24\pm 110.38$&$-0.78\pm 0.05$&$ 17.35\pm 1.03$\\
182&01&J225706.17-002532.8 & 1.985&0380-51792-0274&$  519.08\pm 117.48$&$ 1.05\pm 0.13$&$ 26.30\pm 2.44$\\
182&02&J225706.17-002532.8 & 1.985&0676-52178-0039&$  521.97\pm 131.91$&$ 1.02\pm 0.12$&$ 20.28\pm 2.71$\\
183&01&J231105.54-102837.5 & 3.194&0726-52226-0130&$ 6948.05\pm 210.14$&$-1.13\pm 0.02$&$ 43.91\pm 0.38$\\
183&02&J231105.54-102837.5 & 3.194&0726-52207-0133&$ 7419.36\pm 378.42$&$-1.08\pm 0.03$&$ 43.52\pm 0.58$\\
184&01&J231958.70-002449.3 & 1.890&0382-51816-0064&$ 1329.00\pm 130.74$&$-0.12\pm 0.06$&$ 21.62\pm 1.05$\\
184&02&J231958.70-002449.3 & 1.890&0680-52200-0261&$ 1562.23\pm 130.52$&$-0.72\pm 0.06$&$ 24.83\pm 0.92$\\
185&01&J232840.99-085315.8 & 1.729&0646-52523-0446&$  762.37\pm 100.71$&$-1.47\pm 0.10$&$ 41.52\pm 1.65$\\
185&02&J232840.99-085315.8 & 1.729&3145-54801-0367&$ 1301.68\pm  46.78$&$-1.57\pm 0.03$&$ 40.79\pm 0.47$\\

\hline
\end{tabular}
\end{table*}

\setcounter{table}{1}
\begin{table*}
\centering \caption{-- continue}
\begin{tabular}{lccccccccccccccccccc}

\hline \hline
  No.  & epoch &name  & z& Plate-MJD-Fiber &$L_{1500}$ & $\alpha$ & BAL EW \\
  (1) & (2) & (3) & (4) & (5) & (6) & (7) & (8)\\
\hline

186&01&J234711.44-103742.4 & 1.800&0648-52559-0098&$ 2954.64\pm 116.50$&$-0.07\pm 0.04$&$ 49.98\pm 0.55$\\
186&02&J234711.44-103742.4 & 1.800&2311-54331-0216&$ 2922.86\pm  71.76$&$-0.36\pm 0.03$&$ 51.92\pm 0.41$\\
187&01&J235702.54-004824.0 & 3.013&0387-51791-0246&$ 2437.82\pm 209.11$&$-1.31\pm 0.04$&$ 18.02\pm 1.10$\\
187&02&J235702.54-004824.0 & 3.013&0685-52203-0317&$ 2321.58\pm 213.74$&$-1.38\pm 0.04$&$ 20.08\pm 1.21$\\
188&01&J235859.47-002426.2 & 1.759&0387-51791-0181&$  942.02\pm  96.73$&$-0.98\pm 0.08$&$ 15.85\pm 1.25$\\
188&02&J235859.47-002426.2 & 1.759&0685-52203-0164&$  760.98\pm 105.55$&$-0.43\pm 0.09$&$ 21.50\pm 1.53$\\
\hline
\end{tabular}
\end{table*}

\begin{table*}
\centering \caption{The subsample of 43 two-epoch spectra with identified variable BAL regions. Col. (1) is the No. of the two-epoch spectra. Col. (2) is the name. Col. (3) is the redshift. Col. (4) is one observation of Plate-MJD-Fiber. Col. (5) is another observation of Plate-MJD-Fiber. Col. (6) is the bolometric luminosity in units of \ergs. Col. (7) is the \mgii-based black hole mass in units of $M_{\odot}$. Col. (8) is the Eddington ratio. Col. (9) is $\Delta EW$ of all variable BAL regions in two-epoch different spectra in units of \AA.  Col. (10) is the boundary of all variable BAL regions ($V_{max}, V_{min}$) in units of \kms.}
\label{table3}
\begin{lrbox}{\tablebox}

\begin{tabular}{lccccccccccccccccccc}
\hline \hline
N   &  $name$ & z & Plate-MJD-Fiber$^1$ & Plate-MJD-Fiber$^2$ &$LogL_{bol}$ & $Log M_{BH}$ &$L_{bol}$/$L_{Edd}$ & $\Delta EW^{vary}$ & Vary($V_{max}$, $V_{min}$)\\
(1) & (2) & (3) & (4) & (5) & (6) & (7) & (8) & (9) & (10) \\
\hline
001&J002710.06-094435.3 & 2.070&0653-52145-0556&3105-54825-0310&   46.96 &$ 8.87\pm 0.19$& 0.98&$ -8.45\pm  0.54$&( 27311.7, 22275.5); (  5036.2,  2324.4)\\
002&J003312.25+155442.4 & 1.937&0417-51821-0576&3133-54789-0379&   46.23 &$ 8.86\pm 0.19$& 0.19&$ -4.17\pm  0.59$&( 21113.3, 19951.1); ( 14333.8, 12203.1)\\
003&J004527.68+143816.1 & 1.992&0419-51812-0105&0419-51868-0106&   47.44 &$ 9.70\pm 0.03$& 0.43&$ +0.61\pm  0.17$&(  7360.6,  6198.4)\\
004&J005419.99+002727.9 & 2.490&0394-51876-0514&3111-54800-0509&   47.07 &-- & --&$ -2.76\pm  0.29$&(  8329.1,  6585.8); (  4455.1,  3099.2)\\
005&J010859.53-105757.8 & 1.807&0659-52199-0099&3109-54833-0003&   46.74 &$ 8.93\pm 0.11$& 0.51&$ +5.03\pm  0.74$&( 28861.3, 27311.7); ( 21500.7, 20338.5); ( 11234.6, 10072.4)\\
006&J012603.62-100114.8 & 2.302&0661-52163-0114&2878-54465-0274&   46.93 &-- & --&$ +2.93\pm  0.46$&(  6585.8,  3292.9)\\
007&J020006.31-003709.7 & 2.141&0403-51871-0070&2866-54478-0197&   46.65 &$10.14\pm 0.14$& 0.03&$ +6.68\pm  0.48$&( 20338.5, 18014.1); ( 16851.9, 13365.3); ( 12590.5, 11234.6)\\
008&J021818.14-092153.5 & 1.880&0668-52162-0218&3122-54821-0201&   47.08 &$ 9.80\pm 0.13$& 0.15&$ +0.76\pm  0.21$&( 12009.4, 10847.2)\\
009&J022036.27-081242.9 & 2.004&0668-52162-0547&3122-54821-0482&   46.94 &$ 9.24\pm 0.12$& 0.40&$ -2.07\pm  0.35$&( 27311.7, 26149.5); ( 23825.1, 21888.1)\\
010&J022349.24+004727.8 & 2.335&0704-52205-0459&3127-54835-0338&   46.43 &-- & --&$ +0.89\pm  0.35$&( 11428.3, 10459.8)\\
011&J023252.80-001351.1 & 2.033&0705-52200-0063&3126-54804-0205&   46.70 &$ 9.24\pm 0.19$& 0.23&$ -6.65\pm  0.56$&( 10266.1,  6585.8); (  6198.4,  4261.4)\\
012&J074221.38+165740.3 & 2.538&2074-53437-0207&2915-54497-0216&   46.94 &-- & --&$ -1.20\pm  0.24$&( 23631.4, 22275.5)\\
013&J075007.63+275708.0 & 2.364&1059-52618-0071&2075-53737-0550&   46.94 &-- & --&$ +6.43\pm  0.47$&( 23825.1, 18401.5); ( 10072.4,  6973.2)\\
014&J080455.90+231501.8 & 2.162&1265-52705-0187&1584-52943-0337&   46.79 &$ 9.28\pm 0.23$& 0.26&$ -1.92\pm  0.45$&( 24018.8, 21888.1)\\
015&J081213.95+431715.9 & 1.742&0546-52205-0403&0547-51959-0284&   46.95 &$ 9.37\pm 0.10$& 0.30&$ +4.58\pm  0.60$&( 28086.5, 24212.5)\\
016&J081822.63+434633.8 & 2.042&0547-51959-0122&0547-52207-0157&   46.97 &$ 9.49\pm 0.06$& 0.24&$ +2.95\pm  0.38$&( 13946.4, 10847.2)\\
017&J082238.64+420925.7 & 1.968&0761-52266-0244&0761-54524-0279&   46.56 &$ 9.44\pm 0.12$& 0.10&$ -2.78\pm  0.44$&(  8910.2,  7941.7)\\
018&J084255.92+223431.9 & 2.714&2084-53360-0502&3373-54940-0062&   46.91 &-- & --&$ -4.81\pm  0.60$&( 20338.5, 16270.8); ( 15883.4, 14140.1)\\
019&J092527.71+151416.9 & 1.968&2440-53818-0622&3192-54829-0504&   46.53 &$ 9.05\pm 0.21$& 0.24&$ -4.62\pm  0.52$&( 18788.9, 16077.1); (  9297.6,  7941.7)\\
020&J092720.29+101627.0 & 1.929&1740-53050-0065&3319-54915-0425&   46.74 &$ 9.43\pm 0.06$& 0.16&$ -1.19\pm  0.35$&(  6004.7,  4842.5)\\
021&J093548.50+363121.9 & 2.977&1275-52996-0145&3223-54865-0303&   47.10 &-- & --&$+10.56\pm  0.91$&( 19951.1, 17820.4); ( 16270.8, 11234.6); (  9491.3,  7941.7)\\
022&J094602.23+380059.3 & 2.068&1276-53035-0160&3223-54865-0566&   46.84 &$ 9.38\pm 0.19$& 0.23&$ +2.95\pm  0.38$&( 18207.8, 16270.8);(  9297.6,  8135.4)\\
023&J095901.24+550408.2 & 2.180&0945-52652-0436&3169-54821-0627&   46.79 &$ 9.61\pm 0.12$& 0.12&$ -5.10\pm  0.49$&( 10459.8,  4261.4)\\
024&J100716.69+030438.6 & 2.124&0501-52235-0606&3257-54888-0474&   47.10 &$ 9.42\pm 0.36$& 0.38&$ +4.37\pm  0.28$&( 20725.9, 19176.3); ( 18788.9, 16270.8)\\
025&J101542.04+430455.6 & 2.420&1218-52709-0514&3287-54941-0433&   47.48 &-- & --&$ +5.35\pm  0.24$&(  8716.5,  6198.4); (  5423.6,  3680.3)\\
026&J101616.34+383817.3 & 1.953&1427-52996-0479&3262-54884-0356&   46.38 &$ 9.06\pm 0.21$& 0.17&$ +6.01\pm  0.65$&( 25374.7, 23437.7); ( 21694.4, 19563.7); ( 19176.3, 17820.4)\\
027&J102754.03+182221.6 & 3.078&2591-54140-0453&2868-54451-0518&   47.13 &-- & --&$ +1.10\pm  0.28$&( 13171.6, 11815.7)\\
028&J110208.59+660156.5 & 2.064&0490-51929-0142&3171-54862-0164&   46.59 &$ 9.10\pm 0.25$& 0.25&$ -4.90\pm  0.52$&( 16851.9, 15689.7); ( 15302.3, 12396.8)\\
029&J112239.20+602012.3 & 2.227&0951-52398-0523&3328-54964-0081&   46.66 &$ 9.29\pm 0.15$& 0.19&$ -4.66\pm  0.61$&( 22275.5, 18788.9)\\
030&J112258.77+164540.3 & 3.031&2499-54176-0308&3327-54951-0153&   47.35 &-- & --&$ +1.31\pm  0.20$&( 19757.4, 18401.5); (  8135.4,  7166.9)\\
031&J112703.06+450516.4 & 1.857&1366-53063-0321&3215-54861-0399&   46.67 &$ 9.47\pm 0.14$& 0.13&$ -2.89\pm  0.42$&(  5617.3,  4648.8); (  1937.0,   774.8)\\
032&J120653.39+492919.3 & 1.845&0969-52442-0107&2919-54537-0170&   47.08 &$ 9.65\pm 0.05$& 0.22&$ -1.70\pm  0.25$&( 22662.9, 21500.7); ( 10847.2,  9297.6)\\
033&J120822.25+302423.9 & 2.252&2230-53799-0353&3181-54860-0025&   47.18 &-- & --&$ -1.29\pm  0.28$&( 18014.1, 17045.6); ( 14140.1, 13171.6)\\
034&J121147.38+203402.4 & 2.413&2610-54476-0341&2918-54554-0622&   47.00 &-- & --&$ -6.04\pm  0.52$&( 22081.8, 21113.3); ( 13365.3, 10653.5); ( 10266.1,  7360.6); (  6198.4,  5229.9)\\
035&J122604.28+034317.8 & 1.767&0519-52283-0565&3253-54941-0202&   47.12 &$ 9.63\pm 0.07$& 0.25&$ +1.01\pm  0.20$&( 12009.4, 10653.5)\\
036&J123736.42+143640.1 & 2.704&1768-53442-0038&3254-54889-0548&   46.83 &-- & --&$ +0.98\pm  0.27$&( 22275.5, 21307.0)\\
037&J130136.12+000157.9 & 1.783&0293-51994-0074&0293-51689-0079&   47.16 &$ 9.83\pm 0.08$& 0.17&$ +0.41\pm  0.12$&( 13752.7, 12396.8)\\
038&J131433.83+032321.9 & 2.255&0525-52029-0572&0525-52295-0576&   46.93 &-- & --&$ +3.34\pm  0.45$&(  7360.6,  6004.7); (  4842.5,  3874.0)\\
039&J131505.89+590157.5 & 1.932&0958-52410-0157&3237-54883-0067&   47.10 &$ 9.69\pm 0.05$& 0.20&$ -2.47\pm  0.32$&( 10459.8,  7941.7); (  4067.7,  2905.5)\\
040&J135721.77+005501.1 & 1.997&0301-51942-0408&0301-51641-0411&   47.03 &$ 9.39\pm 0.11$& 0.34&$ -2.94\pm  0.36$&( 26149.5, 23050.3)\\
041&J143117.07+632701.7 & 1.891&0499-51988-0105&2947-54533-0370&   47.17 &$ 9.62\pm 0.06$& 0.28&$ -1.29\pm  0.29$&( 18595.2, 16851.9)\\
042&J150033.52+003353.6 & 2.438&0310-51616-0363&0310-51990-0388&   46.68 &-- & --&$ +7.25\pm  0.58$&( 21500.7, 20338.5); ( 19370.0, 14140.1)\\
043&J232840.99-085315.8 & 1.729&0646-52523-0446&3145-54801-0367&   46.69 &$ 9.43\pm 0.06$& 0.14&$ -2.21\pm  0.47$&( 14721.2, 13365.3)\\
\hline
\end{tabular}
\end{lrbox}
\scalebox{0.70}{\usebox{\tablebox}}
\end{table*}

\end{document}